 \documentclass[final,5p,times,twocolumn]{elsarticle}

\usepackage{graphicx}
\usepackage{amssymb}
\usepackage{amsmath}

\usepackage{lineno}

\biboptions{sort&compress}

\journal{Physics of the Dark Universe}

\begin{document}

\begin{frontmatter}


\title{Sublunar-Mass Primordial Black Holes from Closed Axion Domain Walls}



\author{Shuailiang Ge}

\address{Department of Physics and Astronomy, University of British Columbia, Vancouver, V6T 1Z1, BC, Canada}

\begin{abstract}
We study the formation of primordial black holes (PBHs) from the collapse of closed domain walls (DWs) which naturally arise in QCD axion models near the QCD scale together with the main string-wall network. The size distribution of the closed DWs is determined by percolation theory, from which we further obtain PBH mass distribution and abundance. Various observational constraints on PBH abundance in turn also constrain QCD axion parameter space. Our model prefers axion mass at the meV scale ($f_{a}\sim 10^{9}$ GeV). The corresponding PBHs are in the sublunar-mass window $10^{20}$-$10^{22}$ g (i.e., $10^{-13}$-$10^{-11}M_{\odot}$), one of few mass windows still available for PBHs contributing significantly to dark matter (DM). In our model, PBH abundance could reach $\sim1\%$ of DM, sensitive to the formation efficiency of closed axion DWs. 
\end{abstract}


\end{frontmatter}

\section{Introduction}
\label{sec:introduction}
Primordial black holes (PBHs) have long been considered as viable dark matter (DM) candidates, see Refs.~\cite{sasaki2018primordial,carr2016primordial,Dolgov:2018urv} for recent reviews. Despite various observational constraints, some mass windows remain valid in which PBHs could significantly contribute to DM: sublunar-mass range $\mathcal{O}(10^{20}{\rm g})$ and intermediate mass range $\mathcal{O}(10M_{\odot})$~\cite{sasaki2018primordial,carr2016primordial,carr2018primordial}. In addition to the frequently studied mechanism of PBH formation from the collapse of overdense regions in the early universe~\cite{sasaki2018primordial,carr2016primordial}, PBHs could also be formed from the collapse of topological defects~\cite{hawking1989black,polnarev1991formation,garriga1993black,vilenkin1981cosmological,fort1993global,garriga1993effects,rubin2001formation,khlopov2005primordial,garriga2016black,deng2017primordial}.

QCD axion was originally proposed as a solution to strong CP problem~\cite{peccei1977constraints,weinberg1978new,wilczek1978problem,kim1979weak,shifman1980can,dine1981simple,Zhitnitsky:1980tq}. As Peccei-Quinn (PQ) symmetry gets spontaneously broken at PQ scale $T_{\rm PQ}\sim f_{a}$ in the early universe, axion strings are formed. If PQ symmetry is broken after inflation ($f_{a}\lesssim H_{I}$, post-inflationary scenario), axion domain walls (DWs) will be formed later near QCD scale $T_{1}\sim$ GeV with the pre-existing strings as boundaries, which we call the string-wall network~\cite{vilenkin1982cosmic,sikivie1982axions}. Otherwise, in the pre-inflationary scenario, the pre-existing strings are `blown away' and the axion field gets homogenized by inflation, so no DWs can be formed at $T_{1}$. Propagating axions generated from misalignment mechanism and topological decays are also DM candidates~\cite{sikivie2008axion,marsh2016axion}.

Recently, Refs.~\cite{vachaspati2017lunar,Ferrer:2018uiu} have studied PBH formation from the collapse of closed axion DWs. The PBH mass obtained in Ref.~\cite{vachaspati2017lunar} is $\sim10^{-8}M_{\odot}$ ($10^{25}$ g), but much heavier in Ref.~\cite{Ferrer:2018uiu} $\sim10^{4}$-$10^{7}M_{\odot}$ since an extra bias term is considered there lifting the energy enclosed by DWs. Closed DWs in Refs.~\cite{vachaspati2017lunar,Ferrer:2018uiu} are related to the network fragment which could occur much later than $T_{1}$, and PBH formation there is significantly affected by the fragment time which is however very hard to determine~\cite{hiramatsu2012production, Fleury:2015aca,Klaer:2017ond,Gorghetto:2018myk,Kawasaki:2018bzv}.

In this paper, however, we study the closed axion DWs initially formed at $T_{1}$ together with the main string-wall network. The closed DWs thus evolve independently of the network fragment. Also, we focus on $N_{\rm DW}=1$ case. The size distribution of $N_{\rm DW}=1$ closed DWs initially formed at $T_{1}$ is well predicted by percolation theory, from which we can further calculate the PBH mass distribution and abundance. Another advantage is that $N_{\rm DW}=1$ model naturally avoids the known DW problem that arises in $N_{\rm DW}>1$ models leading to a DW-dominated universe~\cite{sikivie2008axion,Zeldovich:1974uw}. The DW problem in $N_{\rm DW}>1$ cases can also be avoided with a bias term introduced, which is adopted in Ref.~\cite{Ferrer:2018uiu}, although there is only little room in parameter space for this term~\cite{sikivie2008axion}.

In our model, for axion decay constant $f_{a}\sim 10^{9}$ GeV, PBHs formed from the collapse of closed axion DWs are in the sublunar-mass window $\sim10^{20}$-$10^{22}$ g, one of few allowed windows constrained by observations. In addition to the propagating axions generated from misalignment mechanism and topological decays as conventional DM candidates, PBH abundance in our model could reach $\sim1\%$ of DM, sensitive to the formation efficiency of closed DWs at $T_{1}$. Additionally, various observational constraints on PBH abundance in turn could constrain QCD axion parameter space.

The paper is organized as follows. In Section~\ref{sec:size_distribution}, we briefly review the formation of axion DWs and discuss the size distribution of $N_{\rm DW}=1$ closed axion DWs predicted by percolation theory. In Section~\ref{sec:collapse}, we study the criterion for a closed DW to collapse into a black hole. In Section~\ref{sec:DM}, we present the PBH mass distribution and abundance obtained in our model, in comparison with the constraints from astrophysical observations. Also, the constraints on PBH abundance in turn are used to constrain QCD axion parameter space. We draw the conclusions in Section~\ref{sec:conclusion}.

\section{Size distribution of closed axion DWs}
\label{sec:size_distribution}
We start with a brief review of axion DWs formation. Non-perturbative QCD effects induce an effective potential for the axion field $\phi$~\cite{sikivie2008axion,marsh2016axion}: 
\begin{equation}\label{eq:axion_potential}
    V_a=m_{a}^2(T) f_a^2 [1-\cos(\phi/f_a)]
\end{equation}
with $0\leq \phi/f_{a}\leq 2\pi N_{\rm DW}$ where $N_{\rm DW}$ is the model-dependent chiral anomaly coefficient~\cite{peccei2008strong} that also represents the number of degenerate vacua locating at $\phi/f_{a}=2k\pi$. The axion mass is~\cite{Borsanyi:2016ksw,wantz2010axion} 
\begin{equation}\label{eq:axionmass}
    m_{a}(T)= 
\begin{cases}
    f_{a}^{-1}\chi_{0}^{1/2},               &T\leq T_{c}\\
    f_{a}^{-1}\chi_{0}^{1/2}(T/T_{c})^{-\beta},              & T\geq T_{c}
\end{cases}
\end{equation}
where $T_{c}\simeq150$ MeV is the QCD transition temperature, $\chi_{0}=(75.6{\rm~MeV})^4$ is the zero-temperature topological susceptibility and $\beta\simeq4$~\cite{Borsanyi:2016ksw,Gorghetto:2018ocs}.

$V_{a}$ is unimportant until $m_{a}(T)$ increases to the scale of the inverse of Hubble radius $H\sim t^{-1}$ at $t_{1}$~\cite{sikivie2008axion}
\begin{equation}\label{eq:ma1t1}
    m_{a}(t_1) t_1\simeq 1.
\end{equation}
We say axion mass effectively turns on at $t_{1}$. The corresponding temperature is $T_{1}\sim1$ GeV, much lower than PQ scale. In the post-inflationary scenario, axion DWs start to form due to Kibble-Zurek mechanism~\cite{Kibble:1976sj,Zurek:1985qw} at $T_{1}$ when different regions of the universe fall into different vacua. The typical length of each region is the correlation length $\xi$ (see e.g.~Refs.~\cite{Liang:2016tqc,vachaspati2006kinks}):
\begin{equation}\label{eq:xi}
    \xi(T)\simeq m_{a}^{-1}(T)
\end{equation}
Using Eq.~(\ref{eq:ma1t1}), we further get $\xi(T_{1})\simeq t_1$, i.e. the correlation length at DW formation point $t_{1}$ is approximately the Hubble radius.

If $N_{\rm DW}=1$, the topology of vacuum manifold has two discrete values, $\phi/f_{a}=0,2\pi$, corresponding to the same physical vacuum. It is known that DWs can be formed in this case as $\phi$ interpolates between the two topological branches $0$ and $2\pi$~\cite{sikivie2008axion,Vilenkin:2000jqa}, and they could live long enough against tunnelling process to have important implications~\cite{Vilenkin:2000jqa,forbes2001domain}. If we ignore the pre-existing strings at $T_{1}$ (the effects of which will be discussed later), $N_{\rm DW}=1$ model can be treated as $Z_{2}$ model, for they have identical topology of vacuum manifold: both have two discrete values~\cite{Ge:2019voa}. The formation of such walls in the early universe has been widely studied in the literature (see e.g. Refs.~\cite{vachaspati1984formation,harvey1982calculation}): different `cells' (typical length $\xi$) fall into one of the two values randomly with equal probability. Two or more neighbouring cells falling into the same value form a finite cluster (closed DW). A mathematical theory known as percolation theory studies the size distribution of such clusters, which gives~\cite{vachaspati1984formation}: 
\begin{equation}\label{eq:ns}
    n_{s}\propto s^{-\tau}\exp{(-\lambda s^{2/3})}.
\end{equation}
$n_{s}$ is the number density of finite clusters with size $s$ (number of cells within a cluster). $\tau=-1/9$ and $\lambda\approx0.025$ are two coefficients from percolation theory~\footnote{$\lambda$ is obtained indirectly. In percolation theory, $\lambda^{-1}$ is the crossover size where $\lambda^{-1}\simeq \left|p-p_{c}\right|^{\rm -1/\sigma}$ valid for $\left|p-p_{c}\right|\ll 1$ (see e.g. Refs.~\cite{STAUFFER19791,isichenko1992percolation,grinchuk2002large}). $p$ is the probability of each cell choosing one of the two topological branches, so $p=0.5$ in our case; $p_{c}=0.31$ for cubic lattice and $\sigma=0.45$ in 3D~\cite{stauffer2014introduction}, so $\lambda\approx 0.025$ for $\left|p-p_{c}\right|\ll 1$ well satisfied. The other coefficient $\tau=-1/9$ for $p>p_{c}$ is obtained in a field theoretical formulation of the percolation problem~\cite{lubensky1981cluster,stauffer2014introduction}.}. Although Eq.~(\ref{eq:ns}) is originally obtained with the assumption $s\gg 1$, it can be extrapolated down to the smallest clusters $s=1$ with high accuracy~\cite{BAUCHSPIESS1978567}.

Eq.~(\ref{eq:ns}) can be translated into DW language straightforwardly. Finite clusters are closed DWs with volume $R_{1}^3\simeq s\xi^{3}$, where $R_{1}$ is introduced as the radius of closed DWs.
We can write $n_{s}$ in differential form as $n_{s}=d\mathfrak{n}/ds$ where $\mathfrak{n}$ denotes the number density of finite clusters with size \textit{smaller} than $s$. Then, Eq.~(\ref{eq:ns}) becomes 
\begin{equation}\label{eq:fr}
    f(r_{1})=f_{0}\cdot r_{1}^{2-3\tau}\cdot {\rm e}^{\lambda(1-r_{1}^2)}
\end{equation}
where $r_{1}\equiv R_{1}/\xi$, $f(r_{1})\equiv d\mathfrak{n}/dr$. $f_{0}\equiv f(r_{1}=1)$ is the distribution at the smallest size $R_{1}=\xi$. 

Closed DWs are indeed observed in computer simulations. In $Z_2$-system, closed DWs account for $\gamma\sim 13\%$ of total wall area~\cite{vachaspati1984formation}. We expect the proportion is lower in $N_{\rm DW}=1$ models with strings present, because the presence of strings makes less space available to form closed DWs. This has also been seen in simulations~\cite{vachaspati1984formation,Chang:1998tb}. But it's hard to determine the strings effects exactly. One difficulty is that simulations are sensitive to simulation size~\cite{vachaspati1984formation} and may not be properly applied to the universe at $T_{1}$. Another difficulty is that simulations only apply to DWs formed soon after strings formation~\cite{vachaspati1984formation} which contradicts the realistic case $T_{1}\ll T_{\rm PQ}$. Despite simulation difficulties, we can absorb the strings effects on closed DWs at $T_{1}$ into $\gamma$ (defined as the proportion of closed DWs area in total wall area~\cite{Liang:2016tqc}), implying $\gamma\lesssim 13\%$ with strings present. Additionally, in contrast with the traditional view, $N_{\rm DW}=1$ DWs could also be formed in the pre-inflationary scenario ($f_{a}\gtrsim H_{I}$) based on the argument that different topological branches cannot be separated by inflation~\cite{Zhitnitsky:2002qa,Liang:2016tqc}~\footnote{$N_{\rm DW}=1$ closed axion DWs formed in the pre-inflationary scenario are crucial in Refs.~\cite{Zhitnitsky:2002qa,Liang:2016tqc}. The closed walls there accumulate baryons or anti-baryons inside. They finally evolve into the axion quark nuggets (AQNs) which have many intriguing astrophysical and cosmological implications. See the original paper~\cite{Zhitnitsky:2002qa} and recent developments~\cite{Liang:2016tqc,Ge:2017ttc,Ge:2017_2,Zhitnitsky:2017rop,Lawson:2018qkc,Raza:2018gpb,Fischer:2018niu,vanWaerbeke:2018nyj,Liang:2018ecs,Flambaum:2018ohm,Ge:2019voa,Lawson:2019cvy} for details.}. In that scenario, the pre-existing strings are blown away by inflation, so they cannot affect the formation of closed DWs at $T_{1}$, implying that $\gamma\sim13\%$, the same as $Z_{2}$ case.

We can also interpret the correlation length $\xi$ as the average distance among DWs, to get
\begin{equation}\label{eq:wallarea}
    \int_{1}^{\infty}dr_{1}~4\pi(\xi r_{1})^2 f(r_{1})\simeq \gamma\cdot \frac{1}{\xi}.
\end{equation}
The best information we have about $\gamma$ in the post-inflationary scenario is $\gamma\lesssim13\%$ (but nonzero, since closed DWs are observed with strings present~\cite{vachaspati1984formation,Chang:1998tb}). One might worry that closed DWs could be destroyed by intercommuting with walls bounded by strings in the late time evolution \textit{after} $T_{1}$, but our analysis shows that closed DWs will survive, see~\ref{appen:survival} for details. 

\section{Collapse into PBHs}
\label{sec:collapse}
Closed DWs with size $r_{1}>1$ (i.e. $R_{1}>\xi(T_{1})$) are super-Hubble structures since $\xi(T_{1})\simeq t_1$. They do not collapse until the size is surpassed by Hubble horizon. We emphasize that super-Hubble DWs are formed not because $\phi$ is physically correlated in super-Hubble scale, but a natural result of random combinations of self-correlated cells predicted by percolation theory.

Instead of contraction, super-Hubble closed DWs first expand due to the universe's expansion with the scale factor $a(t)\propto T^{-1}\propto t^{1/2}$ (radiation-dominated era). However, the Hubble horizon $H^{-1}\sim t$ increases faster, implying that some time after $t_{1}$ (labeled as $t_{2}$), $H^{-1}$ will catch up with the closed DWs size, $R_{2}\simeq t_{2}$. $R_{1}$ and $R_{2}$ are connected by the universe's expansion, $R_2/R_1\simeq (t_2/t_1)^{1/2}$. Recalling that $r_{1}\equiv R_{1}/\xi(T_{1})\simeq R_{1}/t_1$, we have
\begin{equation}\label{eq:t2t1}
    t_{2}\simeq r_{1}^2 t_1.
\end{equation}
Closed DWs start to collapse at $t_{2}$ as the DW tension overcomes the universe's expansion.

The collapse of closed DWs is dominated by the axion Lagrangian $\mathcal{L}=1/2(\partial_{\mu}\phi)^2-V_{a}$ with $V_{a}$ from Eq.~(\ref{eq:axion_potential}). The equation of motion (EoM) is
\begin{equation}\label{eq:eom}
 \left[\partial^{2}_{t}+\frac{3\partial_{t}}{2t}- \frac{\partial^{2}_{\mathcal{R}}}{a^2(t)}-\frac{2\partial_{\mathcal{R}}}{a^2(t) \mathcal{R}} \right] \tilde{\phi} +m_{a}^2(t)\sin \tilde{\phi}=0
\end{equation}
where we have incorporated the universe's expansion. $\mathcal{R}=R/a(t)$ is the co-moving distance. Also, the axion field is redefined as $\tilde{\phi}=\phi/f_{a}$ (dimensionless). For simplicity, we treat closed DWs as nearly spherical, so the EoM is written in the spherically symmetric form. We can use the kink-antikink pair as the initial configuration of spherical DWs~\cite{vachaspati2017lunar,vachaspati2006kinks}
\begin{equation}\label{eq:wall_initial}
\begin{aligned}
  \tilde{\phi}(t=t_2,\mathcal{R})=&4\left\{ \tan^{-1}[{\rm e}^{m_{a}(t_{2})(\mathcal{R}-R_{2})}]\right. \\
  &+\left.\tan^{-1}[{\rm e}^{m_{a}(t_{2})(-\mathcal{R}-R_{2})}] \right\}
\end{aligned}
\end{equation}
where the initial scale factor is set as $a(t_{2})=1$. We also assume walls initially at rest, $\dot{\tilde{\phi}}(t=t_2,\mathcal{R})=0$. 

Following the procedure of Ref.~\cite{vachaspati2017lunar}, we define $E(t,R)$ as the energy contained within a sphere of radius $R$ at time $t$ during collapse of a closed DW. If for some $t$ and $R$, we have $R$ smaller than the corresponding Schwarzschild radius $R_{s}=2GE(t,R)$, a black hole will be formed. The above criterion can be expressed as~\cite{vachaspati2017lunar}
\begin{equation}\label{eq:cri_0}
    \frac{R_{s}}{R}=\frac{2GE(t,R)}{R}\gtrsim 1~~
    \Rightarrow~~ S(t,R)\gtrsim m_{\rm P}^2  
\end{equation}
where $S(t,R)\equiv 2E(t,R)/R$ and $m_{\rm P}$ is the Planck mass. By numerically solving the EoM (\ref{eq:eom}) with the initial conditions above, we can obtain the evolution of $S(t,R)$. The detailed numerical calculations are shown in~\ref{appen:numerical}. The key result is that the maximum $S(t,R)$ is related to the initial collapse size $R_{2}$ by
\begin{equation}\label{eq:smax}
    S_{\rm max}=k_1[m_{a}(t_2) R_{2}]^{k_2}\cdot f_{a}^2
\end{equation}
where $k_1\approx 3.1\times10^{3}$ and $k_2\approx 2.76$. This should be compared with a similar relation in Ref.~\cite{vachaspati2017lunar} where $k_1\approx 21.9$ and $k_2\approx 2.7$. The crucial difference is that in our model closed DWs are originally formed at $T_{1}$ together with the main network and the collapse point $T_{2}$ could be earlier than the QCD transition $T_{c}$ (i.e. $T_{c}<T_{2}<T_{1}$), so the full expression of axion mass Eq.~(\ref{eq:axionmass}) where $m_{a}(T)$ increases rapidly with $T$ before $T_{c}$ must be included in solving the EoM~(\ref{eq:eom}). Additionally, our EoM includes the universe's expansion. In comparison, Ref.~\cite{vachaspati2017lunar} considered collapse of fragments from the string-wall network. The fragment process could occur later than $T_{c}$, so $m_{a}$ is treated as a constant there. 

Also, fragments in Ref.~\cite{vachaspati2017lunar} inherit angular momentum from strings motion, which could significantly suppress PBH formation. However, our model does not suffer from this suppression. Closed DWs have no initial angular momentum at $T_{1}$ since they are formed independently of the main network, and the simple assumption of spherical shape guarantees no angular motion later but only radial motion.

\begin{figure}
    \centering
    \includegraphics[width=1\linewidth]{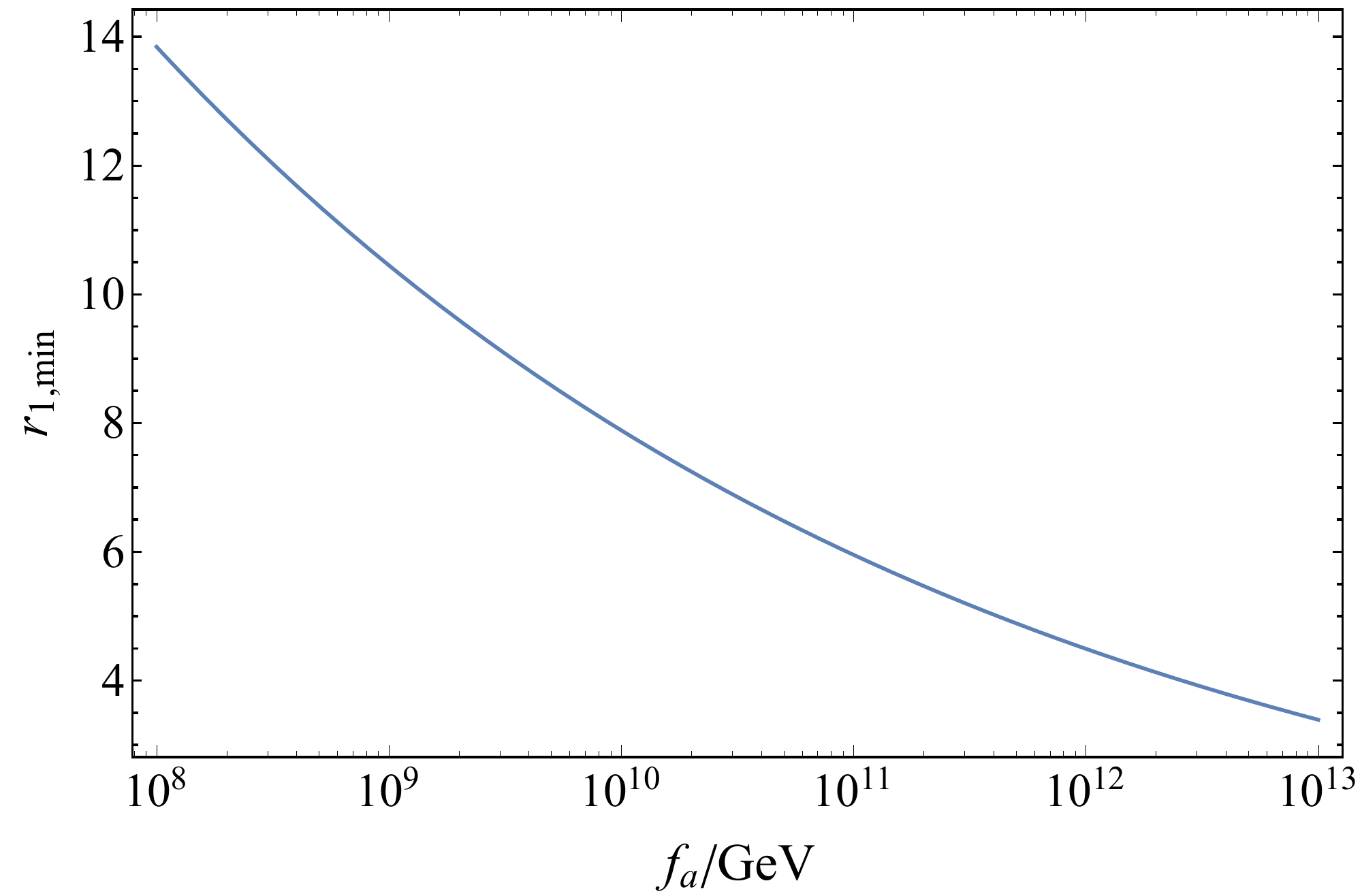}
    \caption{Relation between $r_{\rm 1, min}$ and $f_{a}$.}
    \label{fig:r1fa}
\end{figure}
Substituting Eq.~(\ref{eq:smax}) into Eq.~(\ref{eq:cri_0}) and using Eq.~(\ref{eq:t2t1}), we can finally express the criterion of PBH formation in terms of $r_{1}$: 
\begin{equation}\label{eq:cri_r1}
    r_{1}^2\gtrsim \frac{m_{a}(t_1)}{m_{a}(t_2)}\left(\frac{m_{P}^2}{k_{1}f_{a}^2}\right)^{1/k_{2}}.
\end{equation}
The classical window of current axion mass is $10^{-6}~{\rm eV}\lesssim m_{a,0}\lesssim 10^{-2}~{\rm eV}$~\cite{graham2015experimental}, implying $10^{8}~{\rm GeV}\lesssim f_{a}\lesssim 10^{12}~{\rm GeV}$ [Eq.~(\ref{eq:axionmass})]. $r_{\rm 1, min}$ is the minimum radius satisfying the criterion Eq.~(\ref{eq:cri_r1}). With $f_{a}$ known, $t_{1}$ and $t_{2}$ are also known from Eqs.~(\ref{eq:axionmass}), (\ref{eq:ma1t1}) and (\ref{eq:t2t1}), so $r_{\rm 1, min}$ is merely determined by $f_{a}$. In Fig.~\ref{fig:r1fa}, we plot the relation $r_{\rm1, min}$-$f_{a}$ (see also~\ref{appen:numerical} for more numerical details). 

\section{PBHs as DM}
\label{sec:DM}
Eq.~(\ref{eq:cri_r1}) roughly determines whether a closed axion DW could collapse into a PBH. To exactly calculate the PBH mass, however, we need to answer many complicated questions, e.g. how the PBH as the core alters the wall dynamics and the fraction of the wall falling into the PBH, etc. For simplicity, we estimate the PBH mass as the energy initially stored in the closed wall at $t_{2}$ when it starts to collapse:
\begin{equation}\label{eq:MPBH}
    M_{\rm PBH}\simeq 4\pi R_{2}^2 \sigma(t_{2}) \simeq 4\pi r_{1}^4\cdot m_{a}^{-2}(t_{1})  \cdot \sigma(r_{1}^2 t_{1})
\end{equation}
where $\sigma=8f_{a}^2 m_{a}$ is the DW tension~\cite{Vilenkin:2000jqa}.

\begin{figure}
    \centering
    \includegraphics[width=1\linewidth]{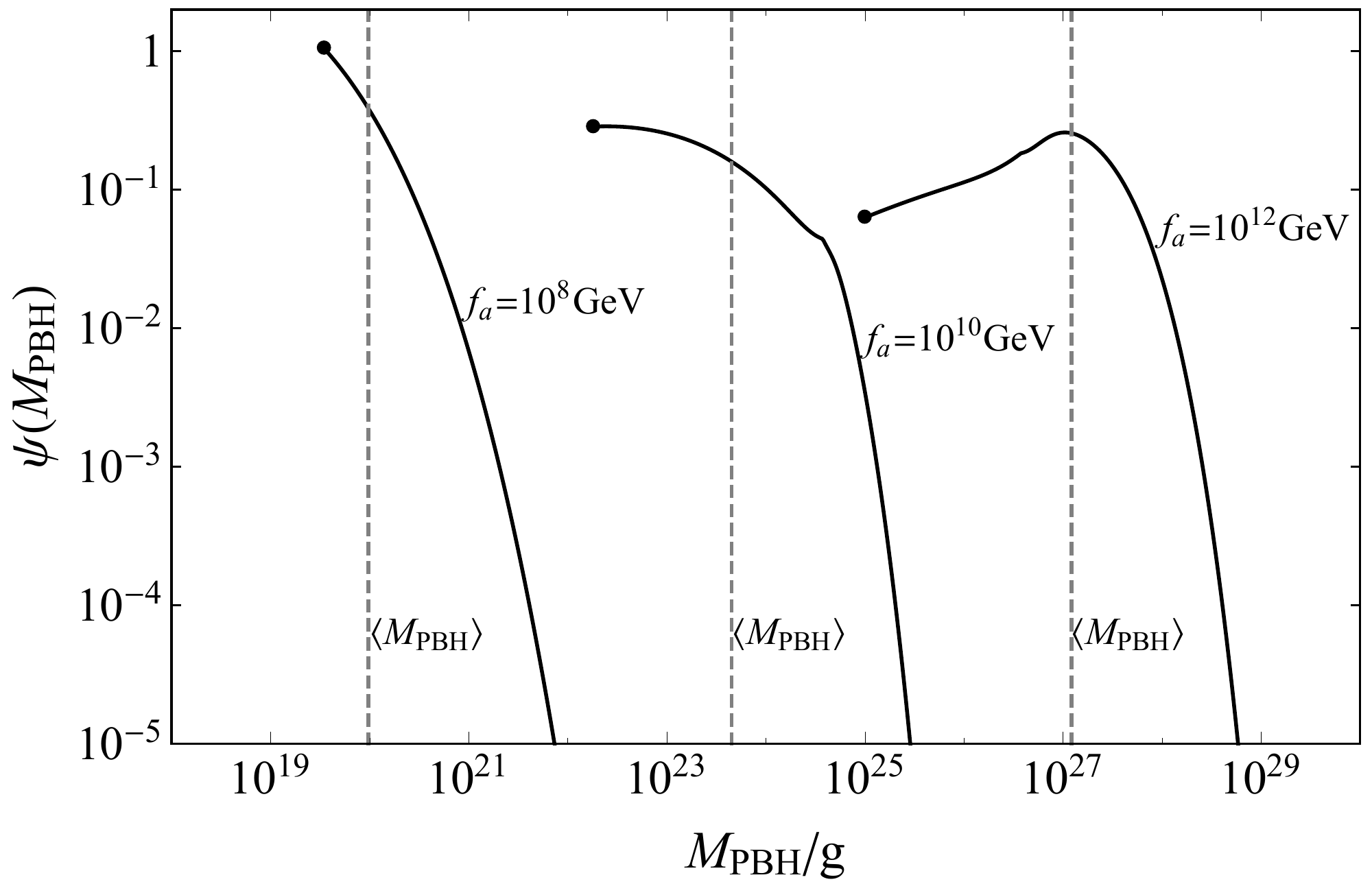}
    \caption{To compare PBH mass distributions for different $f_{a}$, we have rescaled the distribution Eq.~(\ref{eq:massdistribution}) as $\psi(M_{\rm PBH})\equiv (\left<M_{\rm PBH}\right>/\Omega_{\rm PBH}) \cdot (d\Omega_{\rm PBH}/dM_{\rm PBH})$ which is normalized as $\int dM_{\rm PBH}~\psi(M_{\rm PBH})=\left<M_{\rm PBH}\right>$. The black dot and the dashed line for each $f_{a}$ are respectively the minimum PBH mass $M_{\rm PBH, min}$ (corresponding to $r_{\rm1,min}$) and the average mass $\left<M_{\rm PBH}\right>$ Eq.~(\ref{eq:mass_average}).}
    \label{fig:psi}
\end{figure}

The PBH mass distribution is related to the size distribution of closed axion DWs Eq.~(\ref{eq:fr}) via
\begin{equation}
    \frac{d\rho_{\rm PBH}(t)}{dM_{\rm PBH}}=M_{\rm PBH}(r_{1})\cdot f(r_{1}) \cdot \left[\frac{T(t)}{T_{1}}\right]^3 \cdot\frac{d{r_{1}}}{dM_{\rm PBH}}
\end{equation}
where $\rho_{\rm PBH}(t)$ is the mass density of PBHs. $[T(t)/T_{1}]^3$ is the matter density decrease with the universe expanding. We further define $\Omega_{\rm PBH}(t)=\rho_{\rm PBH}(t)/\rho_{\rm cr}(t)$ where $\rho_{\rm cr}(t)=3H^2(t)/8\pi G$ is the critical density. $\Omega_{\rm PBH}(t)$ remains constant after the epoch of matter-radiation equality $T_{\rm eq}\approx 0.8$ eV, so the present mass distribution of PBHs is
\begin{equation}\label{eq:massdistribution}
\frac{d\Omega_{\rm PBH}(t_{\rm eq})}{dM_{\rm PBH}}=\frac{M_{\rm PBH}(r_{1})\cdot f(r_{1})}{\rho_{\rm cr}(t_1)}  \cdot \frac{T_1}{T_{\rm eq}} \cdot\frac{d{r_{1}}}{dM_{\rm PBH}}
\end{equation}
By integrating Eq.~(\ref{eq:massdistribution}), the present PBH abundance is
\begin{equation}\label{eq:Omega}
\Omega_{\rm PBH}=\int_{r_{\rm 1, min}}^{\infty}~\left(\frac{M_{\rm PBH}(r_{1})\cdot f(r_{1})}{\rho_{\rm cr}(t_1)}  \cdot \frac{T_1}{T_{\rm eq}} \right) ~d{r_{1}}.
\end{equation}
The average mass of PBHs can be calculated as
\begin{equation}\label{eq:mass_average}
    \left<M_{\rm PBH}\right>=\frac{\int_{r_{\rm 1,min}}^{\infty}dr_{1}~M_{\rm PBH}(r_{1})f(r_{1})}{\int_{r_{\rm 1,min}}^{\infty}dr_{1}~f(r_{1})},
\end{equation}
which does not change with the universe's expansion. There is a one-to-one correspondence between $\left<M_{\rm PBH}\right>$ and $f_{a}$. In Fig.~\ref{fig:psi}, we plot PBH mass distributions for different $f_{a}$. We see that PBHs are generally within the mass range $10^{19}$-$10^{29}$ g, but the distribution for each $f_{a}$ is quite narrow centering at $\sim\left<M_{\rm PBH}\right>$ and heavy PBHs are greatly suppressed due to Eq.~$(\ref{eq:fr})$. 

We emphasize that PBH mass reaching the scale $10^{19}$-$10^{29}$ g is due to the large size of closed DWs which is inversely proportional to the axion mass at $T_{1}\sim$ GeV, i.e. $\xi\simeq m_{a}^{-1}(T_{1})$, rather than the current axion mass $m_{a,0}$. There is a huge difference between $m_{a,0}$ and $m_{a}(T_{1})$. For example, for $m_{a,0}$ as large as $10^{-4}$ eV, we have $m_{a}(T_{1})\sim 10^{-8}$ eV [Eq.~(\ref{eq:axionmass})]. Another factor contributing to closed DWs size is $r_{1}$ predicted by percolation theory. See also Eq.~(\ref{eq:MPBH}) where $m_{a}^{-1}(T_{1})$ and $r_{1}$ enter the PBH mass expression. 

PBHs surviving today contribute to DM with the trivial constraint $\Omega_{\rm PBH}\leq\Omega_{\rm DW}$. Furthermore, various astrophysical observations constrain $\Omega_{\rm PBH}$ for a wide mass window~\cite{sasaki2018primordial,carr2016primordial}. Most of the valid constraints assume the PBH mass function is monochromatic. Although PBHs in our model have a mass distribution, it is narrow as we see in Fig.~\ref{fig:psi}. If we approximate our model as one which has the monochromatic mass function $M_{\rm PBH}=\left<M_{\rm PBH}\right>$ with the same abundance $\Omega_{\rm PBH}$, the astrophysical constraints on $\Omega_{\rm PBH}$ can be roughly applied to our model.

\begin{figure}
    \centering
    \includegraphics[width=1\linewidth]{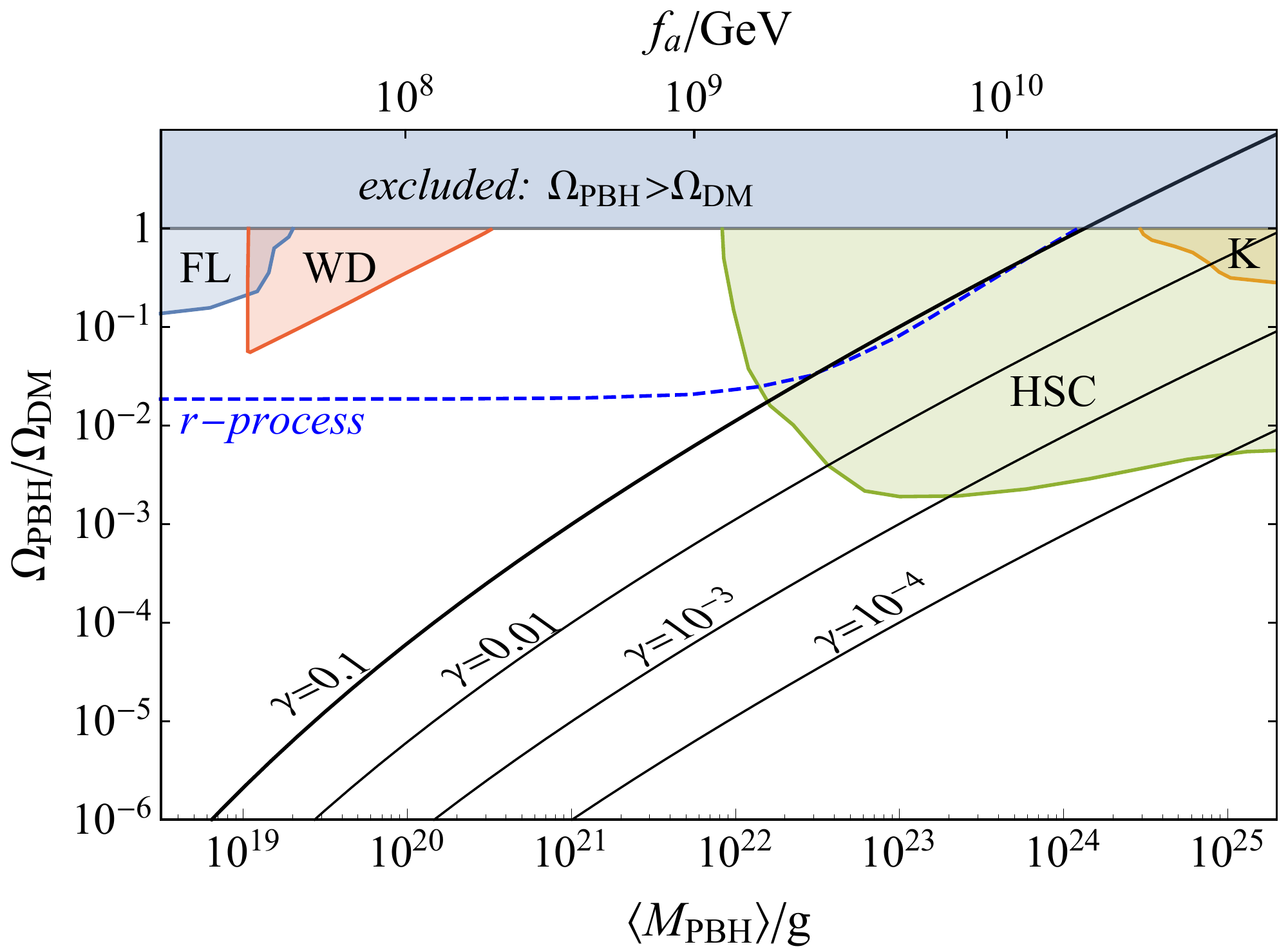}
    \caption{$\Omega_{\rm PBH}/\Omega_{\rm DM}$ as a function of $\left<M_{\rm PBH}\right>$ for various $\gamma$, denoted as black lines. We also plot $f_{a}$-scale in the upper x-axis one-to-one corresponding to $\left<M_{\rm PBH}\right>$. The shaded regions are various observational constraints on PBH abundance: femtolensing (FL)~\cite{femtolensing}, white dwarfs distribution (WD)~\cite{WD}, Subaru/HSC microlensing (HSC)~\cite{HSCmicrolensing} and Kepler microlensing (K)~\cite{Keplermicrolensing}. The r-process nucleosynthesis line is from Ref.~\cite{fuller2017primordial}.}
    \label{fig:Omega}
\end{figure}

$\Omega_{\rm PBH}$ in Eq.~(\ref{eq:Omega}) depends on $f_{a}$ which determines the DWs formation point $t_{1}$ and also the DW tension $\sigma$. Another parameter that also significantly affects $\Omega_{\rm PBH}$ is $\gamma$ [contained in $f(r_{1})$, via Eqs.~(\ref{eq:fr}), (\ref{eq:wallarea})], $\Omega_{\rm PBH}\propto\gamma$. In Fig.~\ref{fig:Omega}, we plot $\Omega_{\rm PBH}/\Omega_{\rm DM}$, the present fraction of PBHs in DM, as a function of $\left<M_{\rm PBH}\right>$ (or $f_{a}$ in the second x-axis, one-to-one corresponding to $\left<M_{\rm PBH}\right>$) for different $\gamma$, with various observational constraints. We see that for $f_{a}\sim 10^{9}$ GeV, PBHs are in the sublunar-mass window $\left<M_{\rm PBH}\right>\sim10^{20}$-$10^{22}$ g, one of few allowed windows~\footnote{Like many other discussions (e.g. Refs.~\cite{fuller2017primordial,HSCmicrolensing}), Fig.~\ref{fig:Omega} does not include the constraint from observations of neutron stars~\cite{Capela:2013yf} which depends on the controversial assumption of PBHs as DM existing in globular clusters. Many observations disfavor DM existing in such regions, see e.g. Ref.~\cite{lane2009testing}.}. For the typical value $\gamma=0.1$, PBHs could account for up to $\sim1\%$ of DW in this mass window. If closed DWs are formed more efficiently, PBHs could contribute more to DM.

We can in turn constrain QCD axion parameter space using the constraints on $\Omega_{\rm PBH}$. Fig.~\ref{fig:Omega} shows that $f_{a}\gtrsim10^{10}$ GeV is almost excluded, although extremely small $\gamma\lesssim 10^{-3}$ is still plausible resulting in $\Omega_{\rm PBH}\lesssim 10^{-3} \Omega_{\rm DM}$. For $f_{a}\lesssim10^{8}$ GeV, PBH abundance is very tiny ($f_{a}\lesssim10^{8}$ GeV is actually excluded by independent observations of supernovae cooling~\cite{chang2018supernova}). Our model prefers $f_{a}\sim 10^{9}$ GeV corresponding to $m_{a,0}\sim$ meV (see a similar result in Ref.~\cite{Ferrer:2018uiu} but depending on a totally different mechanism). Additionally, PBH formation mechanism suggested in this work can also be applied to axion-like particles (ALPs) where $m_a$ and $f_a$ are not linked. In the ALP case, PBH formation could even be more efficient due to the larger DW sizes since the ALP mass could be lower than $10^{-12}$ eV~\cite{Ringwald:2014vqa}.

\section{Conclusions and discussions}
\label{sec:conclusion}
We have studied PBH formation from the collapse of closed QCD axion DWs naturally arising when axion mass effectively turns on. PBH mass distribution can be obtained from the size distribution of closed DWs predicted by percolation theory. Our model prefers axion mass at the meV scale (several experiments can detect axion in this mass range, see Ref.~\cite{Irastorza:2018dyq} for a review). The resulting PBHs are in the sublunar-mass window $10^{20}$-$10^{22}$ g, one of few allowed windows constrained by observations. PBH abundance in our model could vary a lot and it could reach $\sim1\%$ of DM, where the formation efficiency $\gamma$ of closed DWs plays a key role.

Sublunar-mass PBHs have other significant implications. Ref.~\cite{fuller2017primordial} suggests that their interactions with neutron stars could solve the long-standing puzzle of r-process nucleosynthesis, which might get indirect supports from aLIGO, aVirgo and KAGRA experiments~\cite{Cote:2016vla,TheVirgo:2014hva,Aso:2013eba} in the near future. In Fig.~\ref{fig:Omega}, r-process is denoted as the dashed line, the region above/below which is the parameter space that fully/partially explains r-process observations~\cite{fuller2017primordial}. Ref.~\cite{PhysRevLett.91.021101} discussed the possibility of detecting gravitational waves generated by sublunar-mass PBH binaries. Ref.~\cite{Naderi:2017ite} proposed the
sublunar-mass PBHs detection through the diffractive microlensing of quasars in long wavelengths with sublunar-mass PBHs as lenses, which could also detect the PBH mass distribution. These experiments might support or exclude our proposal of PBH formation.

\section*{Acknowledgments}
The work was initiated in the conference IPA 2018 (Interplay between Particle and Astroparticle Physics) in Cincinnati, USA. I thank IPA organizers for this excellent conference. I also thank Ariel Zhitnitsky for useful comments on the work. This work was supported in part by the National Science and Engineering Research Council of Canada and the Four Year Doctoral Fellowship (4YF) of UBC.

\appendix
\section{Survival of the closed axion DWs in the pre-collapse evolution}
\label{appen:survival}
As we discussed in the main text, closed axion DWs are formed at $T_{1}$ and start to collapse at $T_{2}=T_{1}/r_{1}$ when their sizes are surpassed by the Hubble horizon. The minimum $r_{1}$ required to collapse into PBHs is about $4$ to $14$ for different $f_{a}$ as we see in Fig.~\ref{fig:r1fa} in the main text. The \textit{pre}-collapse evolution refers to the evolution of closed axion DWs from $T_{1}$ to $T_{2}$. During this period, in addition to closed DWs, walls bounded by strings (which we call string-wall objects) are also copiously present in the system (post-inflationary scenario), whose intercommuting with closed DWs might destroy closed DWs~\cite{vilenkin1982cosmic}. In this section, we are going to study how string-wall objects affect closed DWs and demonstrate that closed DWs will survive against these effects.

The string-wall objects are formed at $T_{1}$ as strings become boundaries of walls. They are like pancakes or large walls with holes~\cite{Chang:1998tb}. $T_{1}$ can be obtained from Eqs. (\ref{eq:axionmass}) and (\ref{eq:ma1t1}):
\begin{equation}\label{eq:T_1}
    T_{1}\simeq 1{\rm~GeV}\cdot\left(\frac{10^{12}{\rm~GeV}}{f_{a}}\right)^{1/6}.
\end{equation}
Another critical time is the time when the domain wall tension dominates over that of strings. We denote the time as $t_{\rm w}$, which is defined by~\cite{Chang:1998tb,hiramatsu2012production,Vilenkin:1984ib}
\begin{equation}\label{eq:t_w}
    t_{\rm w}\simeq\mu(t_{\rm w})/\sigma(t_{\rm w}),
\end{equation}
where $\sigma\simeq8f_{a}^2 m_{a}$ is the wall tension and $\mu\simeq \pi f_{a}^2{\rm ln}(f_{a}/m_{a})$ is the energy per unit length of strings~\cite{Chang:1998tb}. Solving Eq.~(\ref{eq:t_w}), we get~\cite{Chang:1998tb} 
\begin{equation}\label{eq:T_w}
    T_{\rm w}\simeq 600 {\rm~MeV} \cdot \left(\frac{10^{12}{\rm~GeV}}{f_{a}}\right)^{1/6}
\end{equation}
which is below $T_{1}$. After $T_{\rm w}$, the dynamics of string-wall objects is dominated by walls, whereas, before $T_{\rm w}$ it is dominated by strings~\cite{Chang:1998tb}. Thus, the evolutions of the string-wall objects are totally different before and after $T_{\rm w}$, so we should should discuss their effects on closed walls separately. 

\textit{Before} $T_{\rm w}$. In this stage, we have $t<\mu(t)/\sigma(t)$ and strings dominate the dynamics of string-wall objects. The evolution of strings in this stage is no qualitatively different from that before $T_{1}$ when walls have not been formed yet~\cite{Vilenkin:1984ib}. The main source of strings is closed loops (or wiggles on long strings) with the typical size $t$~\cite{Chang:1998tb}. These strings move relativistically and are likely to hit closed walls, which will create holes on walls~\cite{vilenkin1982cosmic}. However, the holes that are formed in this stage (before $T_{\rm w}$) will shrink and disappear~\cite{Vilenkin:1984ib}. This is because the force of tension in a string $\sim \mu(t)/t$, is greater than the wall tension $\sigma(t)$, for $t<\mu(t)/\sigma(t)$~\cite{Vilenkin:1984ib}. We thus conclude that although the relativistically moving strings may create holes on walls, these holes will disappear themselves as the tension in a string loop can easily overcome the wall tension in this stage. 

On the other hand, at $T_{1}$, closed walls with string holes on them could also be formed initially with strings present. This is one of the reasons why $\gamma\lesssim13\%$ compared to the case without strings. But as we discussed above, these holes tend to disappear themselves in the initial stage, and thus these holey walls initially formed at $T_{1}$ may become closed, which actually brings $\gamma$ closer to $13\%$. This is another thing we can learn from $t<\mu(t)/\sigma(t)$.

\textit{After} $T_{\rm w}$. The wall tension becomes greater than that of strings. In this stage, if strings hit closed walls and create holes on them, these string holes will inevitably increase in size pulled by the walls, which may significantly decrease the rate of closed walls collapsing into PBHs. However, compared with the first stage, the crucial difference is that the motion of a string after $T_{\rm w}$ is greatly constrained by its own wall originally attached, for the walls dominating the dynamics of the string-wall objects. Also, the string-wall objects will quickly decay into axions~\cite{Chang:1998tb}. As we will see below, string-wall objects cannot reach the nearest closed walls before these string-wall objects totally decay. 

In the first stage (before $T_{\rm w}$), the strings move at relativistic speeds~\cite{Chang:1998tb}. If a string and a wall collide, the intercommuting probability is very high (close to $1$)~\cite{vilenkin1982cosmic,Vilenkin:1984ib,Shellard}. Thus, large closed walls will eat the incoming string-wall objects quickly and efficiently in the first stage (the holes created will disappear as discussed above). With the surrounding regions cleared up, the typical distance between a closed wall surface and the neighbouring string-wall object is the Hubble scale $\sim t$, saturating the requirement of causality\footnote{This is also commonly assumed in many related studies of topological defects where the interactions are efficient, see e.g. Refs.~\cite{sikivie2008axion,Ryden}. This is also consistent with the numerical simulations of string-wall objects where the wall area parameter $\mathcal{A}\lesssim1$~\cite{hiramatsu2012production}, implying on average there is one or less horizon-size string-wall object per horizon.}. The equilibrium will be kept until $T_{\rm w}$ when the dynamics of string-wall objects is greatly altered. Now at $T_{\rm w}$, for string-wall objects, more energy is stored in walls rather than strings and thus the bulk motion of string-wall objects is determined by walls. We should check what will happen to the system. The simulation result of walls speed is $v\sim 0.4c$~\cite{Ryden}. At $T_{\rm w}$, the distance between a string-wall object and its nearest closed wall surface is $\sim t_{\rm w}$. Then, the time needed for the string-wall object to hit the closed wall can be estimated as
\begin{equation}
   \int_{t_{\rm w}}^{t_{\rm hit}}\frac{vdt}{a(t)}\simeq \frac{t_{\rm w}}{a(t_{\rm w})}
\end{equation}
from which we get 
\begin{equation}
    t_{\rm hit}\simeq5.1 t_{\rm w},~~~~T_{\rm hit}\simeq 0.44T_{\rm w}\simeq 0.26 T_{\rm 1},
\end{equation}
To obtain $T_{\rm hit}\simeq 0.26 T_{\rm 1}$, we also used $T_{\rm w}\simeq 0.6T_{1}$ [Eqs.~(\ref{eq:T_1}) and (\ref{eq:T_w})]. 

$T_{\rm hit}$ should be compared with the temperature at which the string-wall objects totally decay. Soon after $T_{\rm w}$, string-wall objects will decay into axions, as the strings pulled by the wall tension quickly unzip the attached walls~\cite{Chang:1998tb}. Recent simulations show that string-wall objects totally decay at $T_{\rm decay}\simeq T_{1}/3$~\cite{Fleury:2015aca}\footnote{
It is $T_{\rm decay}\simeq T_{1}/4$ obtained in Ref.~\cite{Klaer:2017ond}. However, the exact value of $T_{\rm decay}$ is not essential for us. As we will see below, in the realistic case that $m_{a}(t)$ increases rapidly with time, the wall speed is much lower, which finally leads to Eq.~(\ref{eq:Thitll}).}. The crucial point for us is that 
\begin{equation}~\label{eq:ThitTdecay}
    T_{\rm hit}\lesssim T_{\rm decay}
\end{equation}
which implies that string-wall objects cannot reach the nearest closed walls before these string-wall objects totally decay into free axions. In other words, closed domain walls will not be destroyed by the string-wall objects after $T_{\rm w}$.

One more comment is that the wall speed $v\sim 0.4c$ obtained in Refs.~\cite{Ryden} is relatively high, because they did not consider that the axion mass $m_{a}(t)$ increases with time drastically. With the time-dependent $m_{a}(t)$ taken into consideration, the bulk speed is expected to be lower (even non-relativistic). This could be possibly explained as follows. The speed $v$ is related to the ratio of kinetic energy to rest energy $E_{\rm kin}/E_{\rm rest}$~\cite{Ryden,Press} where $E_{\rm kin}\sim\left<\frac{1}{2}\dot{\phi}^2\right> $ and $E_{\rm rest}\sim\left<\frac{1}{2}(\nabla\phi)^2+m_{a}^{2}(t)\right>$. With $m_{a}(t)\propto T^{-\beta}$ increasing rapidly, the ratio becomes much lower and so does the wall speed $v$. We could see this picture more intuitively in Fig.2 of Ref.~\cite{hiramatsu2012production}, where the simulations show that the string-wall objects are constrained ``locally" to decay with almost no bulk motion (close to zero)\footnote{The bulk motion should not be confused with the strings motion pulled by the walls. After $T_{\rm w}$, due to the wall tension, a string is accelerated to relativistic speed in the direction of the wall to which it is originally attached (``unzip")~\cite{Chang:1998tb}. So the strings motion is constrained locally by the position of walls in the string-wall objects (see e.g. Fig.2 of Ref.~\cite{hiramatsu2012production}). However, the bulk speed of the string-wall objects is low as we have discussed.}. Thus, Eq.~(\ref{eq:ThitTdecay}) is quite conservative, and actually we should have 
\begin{equation}\label{eq:Thitll}
T_{\rm hit}\ll T_{\rm decay}.
\end{equation}


We conclude this section that closed walls will survive the pre-collapse evolution. Therefore, $\gamma$ formed at $T_{1}$ remains unaffected and becomes important in calculating the PBH abundance.

\section{Numerical details of the collapse of closed axion DWs}
\label{appen:numerical}
In this section, we are going to show the details of numerically solving the collapse of closed axion DWs, including how we get the expression of $S_{\rm max}$ as shown in Eq.~(\ref{eq:smax}) and also the relation between $r_{\rm 1, min}$ and $f_{a}$ as plotted in Fig.~\ref{fig:r1fa} in the main text.

For the convenience of numerical calculations, we define $\tilde{r}=\mathcal{R}/m_{a}^{-1}(t_{2})$ and $\tilde{t}=t/m_{a}^{-1}(t_{2})$ as dimensionless variables, then the EoM Eq.~(\ref{eq:eom}) and the initial conditions (Eq.~(\ref{eq:wall_initial}) and $\dot{\tilde{\phi}}(t=t_{2},\mathcal{R})=0$) can be written as 
\begin{equation}\label{eq:eom_appen}
 \frac{\partial^2 \tilde{\phi}}{\partial\tilde{t}^2}+\frac{3}{2\tilde{t}}\frac{\partial \tilde{\phi}}{\partial\tilde{t}}-\frac{1}{a^{2}(\tilde{t})}\left(\frac{\partial^{2} \tilde{\phi}}{\partial\tilde{r}^2}+\frac{2}{\tilde{r}}\frac{\partial \tilde{\phi}}{\partial\tilde{r}}\right) +\frac{m_{a}^2(\tilde{t})}{m_{a}^2(\tilde{t}_{2})}\sin \tilde{\phi}=0,
\end{equation}
\begin{equation}\label{eq:wall_initial_appen}
  \tilde{\phi}(\tilde{t_2},\tilde{r})=4\left\{ \tan^{-1}[{\rm e}^{(\tilde{r}-\tilde{r}_{2})}]\right. +\left.\tan^{-1}[{\rm e}^{(-\tilde{r}-\tilde{r}_{2})}] \right\},
\end{equation}
\begin{equation}\label{eq:wall_initial2_appen}
  \left.\frac{\partial \tilde{\phi}(\tilde{t},\tilde{r})}{\partial\tilde{t}}\right|_{\tilde{t}=\tilde{t_2}}=0
\end{equation}
where $\tilde{r_{2}}=R_{2}/m_{a}^{-1}(t_{2})$ and $\tilde{t}_{2}=t_2/m_{a}^{-1}(t_{2})$ are respectively the rescaled initial radius and rescaled initial time at the starting point of the collapse of closed DWs, consistent with the definitions of $\tilde{r}$ and $\tilde{t}$. Note that $\tilde{r}_2=\tilde{t}_2$ since $R_{2}=t_{2}$. As we mentioned in the main text, the initial scale factor is set as $1$, $a(\tilde{t}_2)=1$. In the radiation-dominated era, we have
\begin{equation}\label{eq:at_appen}
    a(\tilde{t})=\left(\frac{t}{t_{2}}\right)^{1/2}=\left(\frac{\tilde{t}}{\tilde{t}_{2}}\right)^{1/2}.
\end{equation}
If PBHs are formed before the QCD transition $T_{c}$, according to Eq.~(\ref{eq:axionmass}) the axion mass that enters Eq.~(\ref{eq:eom_appen}) is 
\begin{equation}\label{eq:ma_appen}
    \frac{m_{a}(\tilde{t})}{m_{a}(\tilde{t}_{2})}=\left(\frac{t}{t_{2}}\right)^{\beta/2}=\left(\frac{\tilde{t}}{\tilde{t}_{2}}\right)^{\beta/2}.
\end{equation}
Later, we will discuss the effect of QCD transition on the collapse of closed axion DWs. As we mentioned in the main text, $\beta\simeq4$. One of the most recent calculations on axion mass is given by Ref.~\cite{Borsanyi:2016ksw} based on lattice QCD method which shows that the exact value is $\beta=3.925$~\footnote{Ref.~\cite{Borsanyi:2016ksw} does not give the value of $\beta$ directly, but the Supplementary Information of that paper provides the related data. By fitting the data provided, we get $\beta=3.925$.}. 

$E(t,R)$ is defined as the energy contained within a sphere of radius $R$ at time $t$ during collapse of a closed DW, which can be calculated as
\begin{equation}
\begin{aligned}
    \frac{E(\tilde{t},\tilde{r})}{f_{a}^2}=&m_{a}^{-1}(\tilde{t}_{2})\cdot\int_{0}^{\tilde{r}} d\tilde{r}'\cdot 4\pi\tilde{r}'^2 \cdot a^{3}(\tilde{t})\cdot \left[\frac{1}{2}\left(\frac{\partial \tilde{\phi}}{\partial\tilde{t}}\right)^2 \right. \\
    &\left.+\frac{1}{2a^{2}(\tilde{t})}\left(\frac{\partial \tilde{\phi}}{\partial\tilde{r}'}\right)^2+\frac{m_{a}^2(\tilde{t})}{m_{a}^2(\tilde{t}_{2})}(1-\cos \tilde{\phi})\right].
\end{aligned}
\end{equation}
We add the prefactor $1/f_{a}^2$ in LHS because $\phi$ is redefined as a dimensionless variable $\tilde{\phi}=\phi/f_{a}$ as we mentioned in the main text. Now, the term $S(t,R)$ related to the criterion of PBH formation can be expressed as 
\begin{equation}\label{eq:s_appen}
    S(\tilde{t},\tilde{r})=\frac{2E(\tilde{t},\tilde{r})}{R}=\frac{2E(\tilde{t},\tilde{r})}{\tilde{r}}\cdot \frac{m_{a}(\tilde{t}_2)}{a(\tilde{t})}.
\end{equation}
The maximum value of $S(\tilde{t},\tilde{r})$ during the collapse is 
\begin{equation}
    S_{\rm max}=\underset{(\tilde{t},\tilde{r})}{\max}~S(\tilde{t},\tilde{r})
\end{equation}
We see that $S_{\rm max}/f_{a}^2$ is a function of $\tilde{r}_2$.

\begin{figure}
    \centering
    \includegraphics[width=1\linewidth]{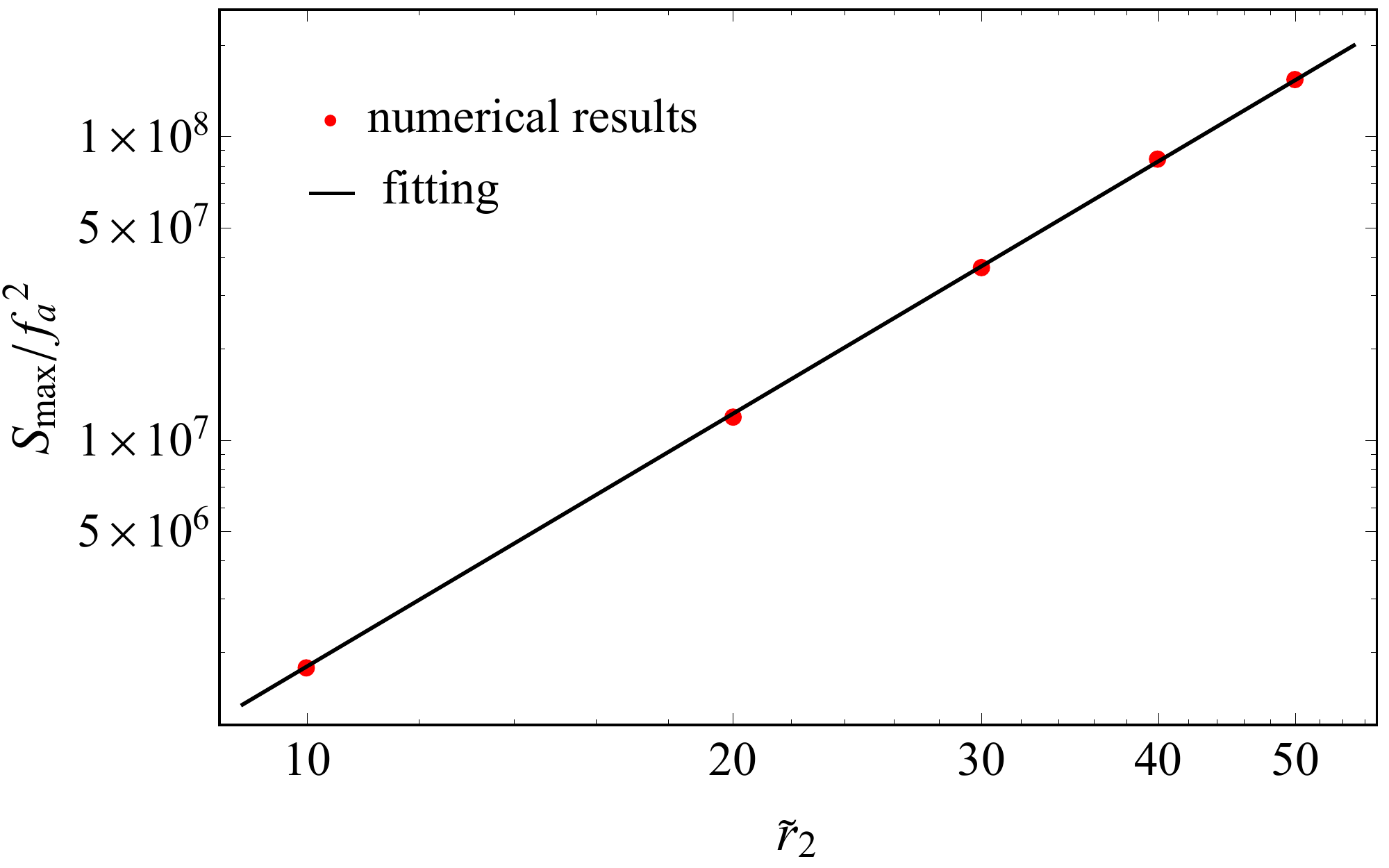}
    \caption{Relation between $S_{\rm max}/f_{a}^2$ and $\tilde{r}_2$. We do numerically for initial radius $\tilde{r}_2=10,20,...,50$ respectively, and the numerical results of $(\tilde{r}_2, S_{\rm max}/f_{a}^2)$ are plotted as red points. The black line is the fitting result Eq.~(\ref{eq:smaxnum_appen}). }
    \label{fig:Smax}
\end{figure}

\begin{figure}
    \centering
    \includegraphics[width=1\linewidth]{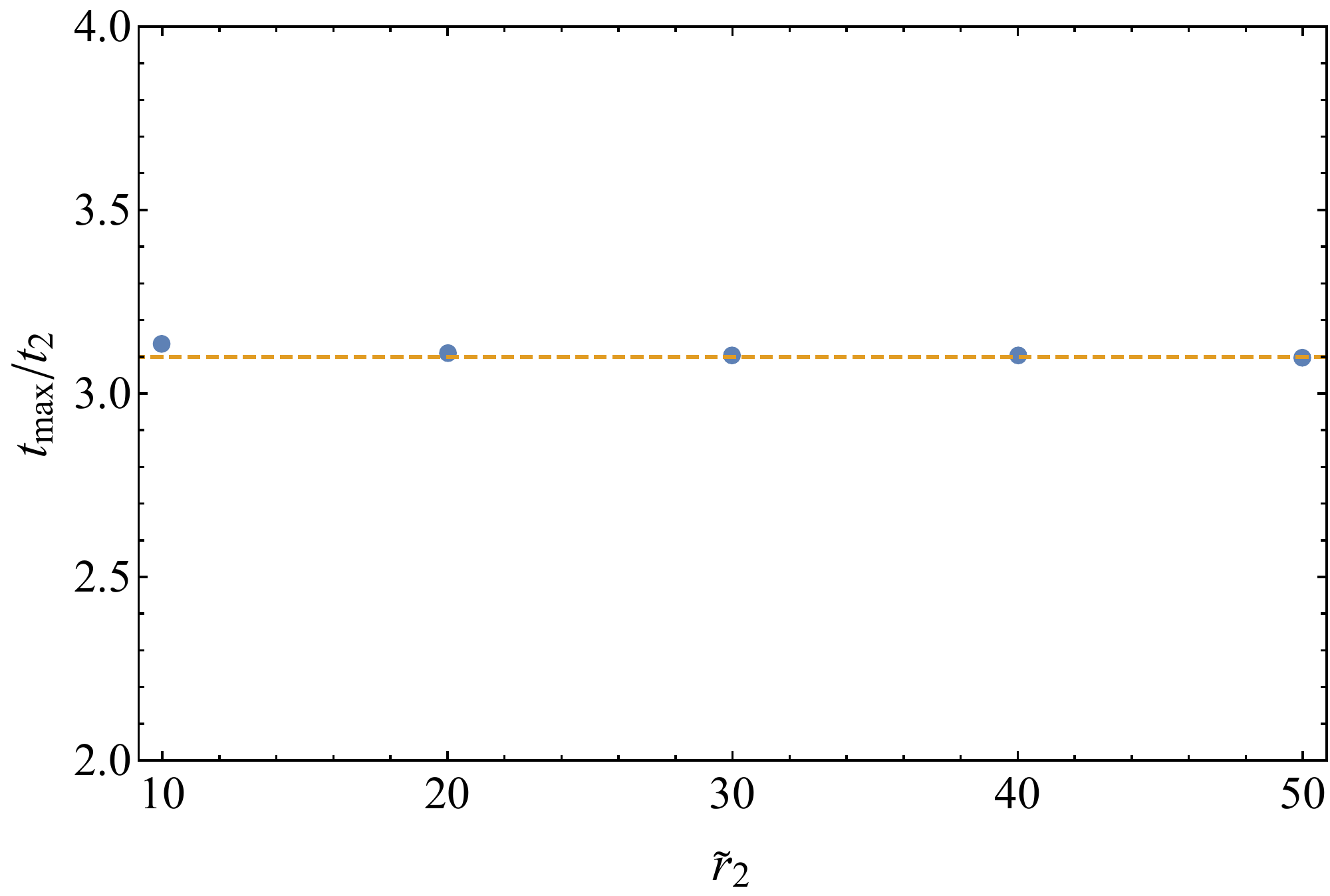}
    \caption{Relation between $t_{\rm max}/t_2$ and $\tilde{r}_2$. The blue points are numerical results and the dashed line is $t_{\rm max}/t_2=3.1$.}
    \label{fig:tmax}
\end{figure}

We then study the collapse of closed axion DWs by numerically solving Eqs.~(\ref{eq:eom_appen})-(\ref{eq:ma_appen}), from which we obtain the evolution of $S(\tilde{t},\tilde{r})$ (based on Eq.~(\ref{eq:s_appen})) and further $S_{\rm max}$. We do numerical calculations for different values of the initial radius $\tilde{r}_2$, and finally we obtain the relation between $S_{\rm max}/f_{a}^2$ and $\tilde{r}_2$ which is plotted in Fig.~\ref{fig:Smax}. We see that $S_{\rm max}/f_{a}^2$ linearly depends on $\tilde{r}_2$ in the log-log scale, consistent with Ref.~\cite{vachaspati2017lunar} which however did the numerical calculations for a constant $m_{a}$. By fitting the numerical results in Fig.~\ref{fig:Smax}, we get 
\begin{equation}\label{eq:smaxnum_appen}
    S_{\rm max}/f_{a}^2= k_{1}\cdot(\tilde{r}_{2})^{k_{2}},
\end{equation}
where $k_{1}=3106.28$ and $k_{2}=2.7626$. In Fig.~\ref{fig:tmax}, we also plot the relation between $t_{\rm max}$ and $\tilde{r}_2$ where $t_{\rm max}$ is the time when $S(\tilde{t},\tilde{r})$ reaches its maximum value $S_{\rm max}$. The numerical results show that 
\begin{equation}\label{eq:tmax_appen}
   t_{\rm max}/t_{2}\approx 3.1. 
\end{equation}
We see that the collapse is a very fast process, with the scale factor $a(t)$ only enlarged by $(t_{\rm max}/t_{2})^{1/2}\approx1.76$ times from $t_{2}$ to $t_{\rm max}$. Similar to Ref.~\cite{vachaspati2017lunar}, we also observed that $S_{\rm max}$ is reached when the wall collapses to the radius close to zero. So the speed of collapse can be estimated as $(t_{\rm max}/t_{2})^{1/2}t_{2}/(t_{\rm max}-t_2)\approx 0.84$, close to the speed of light. 

Substituting Eq.~(\ref{eq:smaxnum_appen}) into the criterion Eq.~(\ref{eq:cri_0}), and using Eqs.~(\ref{eq:ma1t1}) and (\ref{eq:t2t1}), the criterion of PBH formation can be expressed in terms of $r_{1}$:
\begin{equation}\label{eq:cri_r1_appen}
    r_{1}^2\gtrsim \frac{m_{a}(t_1)}{m_{a}(t_2)}\left(\frac{m_{P}^2}{k_{1}f_{a}^2}\right)^{1/k_{2}}.
\end{equation}
Taking equal sign in Eq.~(\ref{eq:cri_r1_appen}), we obtain the lowest limit of the size of closed axion DWs at the formation point $t_{1}$ which could finally collapse into PBHs, denoted as $r_{\rm 1, min}$. 

\begin{figure}
    \centering
    \includegraphics[width=1\linewidth]{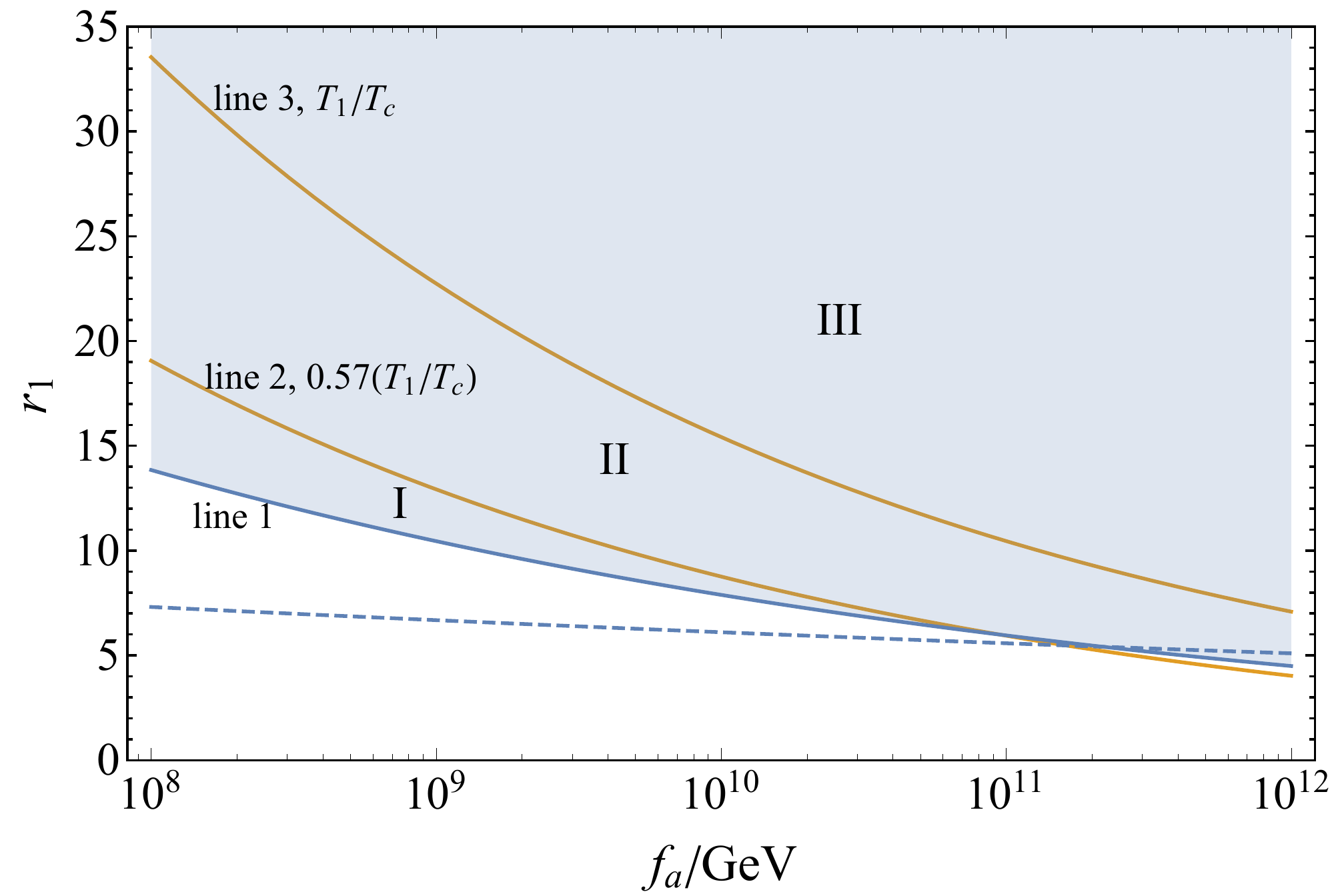}
    \caption{Parameter space $(r_{1}, f_{a})$ for closed axion DWs. Line 1 is $r_{\rm 1, min}$ in Eq.~(\ref{eq:r1min_appen}); the dashed line is $r_{\rm 1, min}$ in Eq.~(\ref{eq:r1min2_appen}). Line 2 and line 3 are respectively the value of $0.57(T_{1}/T_{c})$ and $T_{1}/T_c$ as a function of $f_{a}$. }
    \label{fig:cri}
\end{figure}

However, Eq.~(\ref{eq:smaxnum_appen}) is only applicable when the axion mass relation Eq.~(\ref{eq:ma_appen}) works, which assumes that $S_{\rm max}$ is reached before QCD transition, i.e. $t_{\rm max}< t_{c}$. Using Eqs.~(\ref{eq:t2t1}) and (\ref{eq:tmax_appen}), this condition ($t_{\rm max}< t_{c}$) becomes a constraint on the size of closed DWs at the formation point: 
\begin{equation}\label{eq:condition_appen}
    r_{1}< 0.57 \frac{T_{1}}{T_{c}}.
\end{equation}
The interpretation of this relation is straightforward. The larger a closed DW is at $t_{1}$, the later it will collapse according to Eq.~(\ref{eq:t2t1}), so a sufficiently large closed DW will collapse after $T_{c}\simeq 150$ MeV. If Eq.~(\ref{eq:condition_appen}) is satisfied, we can substitute the axion mass relation Eq.~(\ref{eq:ma_appen}) into Eq.~(\ref{eq:cri_r1_appen}) to get
\begin{equation}\label{eq:r1min_appen}
    r_{\rm 1,min}\simeq \left(\frac{m_{P}^2}{k_{1}f_{a}^2}\right)^{\frac{1}{k_{2}}\cdot\frac{1}{\beta+2}},~{\rm for}~t_{\rm max}<t_{c}.
\end{equation}
We see that $r_{\rm 1,min}$ is merely determined by $f_{a}$. The relation between $r_{\rm1, min}$ and $f_{a}$ is plotted in Fig.~\ref{fig:cri}, denoted as line 1.

For the case $t_{2}>t_{c}$, i.e. closed axion DWs start to collapse after QCD transition, the axion mass that enters the EoM is a constant according to Eq.~(\ref{eq:axionmass}). $t_{2}>t_{c}$ corresponds to the condition $r_{1}>T_1/T_c$. Ref.~\cite{vachaspati2017lunar} numerically solves the collapse of closed axion DWs with $m_{a}$ constant, in which $S_{\rm max}$ has the same form as Eq.~(\ref{eq:smaxnum_appen}) but with $k_{1}\approx21.9$ and $k_{2}\approx2.7$~\footnote{Although Ref.~\cite{vachaspati2017lunar} does not incorporate the effect of the universe's expansion into the EoM, the results of that paper can still be applied here for constant axion mass. This is because the universe's expansion plays only a minor role as we see in Eq.~(\ref{eq:tmax_appen}) where the scale factor is only enlarged by $1.76$ times during the collapse which is a very fast process.}. Then, from Eq.~(\ref{eq:cri_r1_appen}) we can derive $r_{\rm1, min}$ in this case:
\begin{equation}\label{eq:r1min2_appen}
    r_{\rm 1, min}\simeq \left[\frac{m_{a}(t_1)}{m_{a,0}}\right]^{\frac{1}{2}}\left(\frac{m_{P}^2}{21.9f_{a}^2}\right)^{\frac{1}{2.7} \cdot \frac{1}{2}},~{\rm for}~ t_{2}>t_{c}.
\end{equation}
We also plot $r_{\rm 1, min}$ in this case as a function of $f_{a}$ in Fig.~\ref{fig:cri}, denoted as the dashed line.

In Fig.~\ref{fig:cri}, we also plot $T_{1}/T_{c}$ and $0.57(T_{1}/T_{c})$ in comparison with Eqs.~(\ref{eq:r1min_appen}) and (\ref{eq:r1min2_appen}). Region I (between line 1 and line 2) is the parameter space where the condition Eq.~(\ref{eq:condition_appen}) is satisfied, so the criterion Eq.~(\ref{eq:r1min_appen}) is applicable here and the closed DWs with parameters in this region will finally collapse into PBHs. Region III (beyond line 3) is the parameter space where $r_{1}>T_{1}/T_{c}$ (i.e. $t_2>t_{c}$), so we should use the criterion Eq.~(\ref{eq:r1min2_appen}) here. We see that region III is well above the criterion Eq.~(\ref{eq:r1min2_appen}), so the closed DWs with parameters in this region will finally collapse into PBHs. Region II (between line 2 and line 3) where $0.57(T_{1}/T_{c})<r_{1}<T_{1}/T_{c}$ is more subtle. The collapse of closed DWs with parameters in this region will pass through QCD transition, i.e. experience the `knee' of axion mass expression Eq.~(\ref{eq:axionmass}). Since region II satisfies well the criterion of PBH formation from the perspective of both the changing axion mass (Eq.~(\ref{eq:r1min_appen})) and the constant axion mass (Eq.~(\ref{eq:r1min2_appen})), we should expect the closed DWs with parameters in this region will collapse into PBHs~\footnote{\label{foot:1}One may notice that in Fig.~\ref{fig:cri}, the lower three lines (line 1, 2 and the dashed line) intersect with one another at $f_{a}\gtrsim 10^{11}$ GeV and are thus not in good order, which might slightly affect $r_{\rm 1, min}$ in the range $f_{a}\gtrsim 10^{11}$ GeV. However, we may safely ignore the tiny difference since the three lines are very close to each other in this range of $f_{a}$. Also, as we discussed in the main text, the parameter space $f_{a}\gtrsim 10^{11}$ GeV is less interesting since it is almost excluded by observational constraints on $\Omega_{\rm PBH}$. The most interesting part is $f_{a}\sim 10^{9}$ GeV which results in sublunar-mass PBHs, and $r_{\rm 1, min}$ can be well determined for $f_{a}\lesssim 10^{11}$ GeV as we see in Fig.~\ref{fig:cri}.}. 

To conclude, region I, II, and III are all parameter spaces (the shaded region) where closed axion DWs can collapse into PBHs. Thus, the criterion Eq.~(\ref{eq:r1min_appen}) denoted as line 1 in Fig.~\ref{fig:cri} is indeed the lowest limit of $r_{1}$ for PBH formation (the tiny difference in the range $f_{a}\gtrsim10^{11}$ GeV can be ignored as we discussed in footnote~\ref{foot:1}), which is also plotted in Fig.~\ref{fig:r1fa} in the main text. Note that we cannot use Eq.~(\ref{eq:r1min2_appen}) (dashed line) as the final criterion although it is lower than line 1, because the parameter space around the dashed line satisfies the condition Eq.~(\ref{eq:condition_appen}) and thus should be checked by the criterion Eq.~(\ref{eq:r1min_appen}) rather than Eq.~(\ref{eq:r1min2_appen}).

\bibliographystyle{model1-num-names}
\bibliography{main}

\begin{thebibliography}{84}
\expandafter\ifx\csname natexlab\endcsname\relax\def\natexlab#1{#1}\fi
\providecommand{\bibinfo}[2]{#2}
\ifx\xfnm\relax \def\xfnm[#1]{\unskip,\space#1}\fi
\bibitem[{Sasaki et~al.(2018)Sasaki, Suyama, Tanaka, and
  Yokoyama}]{sasaki2018primordial}
\bibinfo{author}{M.~Sasaki}, \bibinfo{author}{T.~Suyama},
  \bibinfo{author}{T.~Tanaka}, \bibinfo{author}{S.~Yokoyama},
\newblock \bibinfo{title}{Primordial black holes—perspectives in
  gravitational wave astronomy},
\newblock \bibinfo{journal}{Classical and Quantum Gravity} \bibinfo{volume}{35}
  (\bibinfo{year}{2018}) \bibinfo{pages}{063001}.
\bibitem[{Carr et~al.(2016)Carr, K{\"u}hnel, and Sandstad}]{carr2016primordial}
\bibinfo{author}{B.~Carr}, \bibinfo{author}{F.~K{\"u}hnel},
  \bibinfo{author}{M.~Sandstad},
\newblock \bibinfo{title}{Primordial black holes as dark matter},
\newblock \bibinfo{journal}{Physical Review D} \bibinfo{volume}{94}
  (\bibinfo{year}{2016}) \bibinfo{pages}{083504}.
\bibitem[{Dolgov(2018)}]{Dolgov:2018urv}
\bibinfo{author}{A.~D. Dolgov},
\newblock \bibinfo{title}{{Massive Primordial Black Holes in Contemporary and
  Young Universe (old predictions and new data)}},
\newblock \bibinfo{journal}{Int. J. Mod. Phys.} \bibinfo{volume}{A33}
  (\bibinfo{year}{2018}) \bibinfo{pages}{1844029}.
\bibitem[{Carr and Silk(2018)}]{carr2018primordial}
\bibinfo{author}{B.~Carr}, \bibinfo{author}{J.~Silk},
\newblock \bibinfo{title}{Primordial black holes as generators of cosmic
  structures},
\newblock \bibinfo{journal}{Monthly Notices of the Royal Astronomical Society}
  \bibinfo{volume}{478} (\bibinfo{year}{2018}) \bibinfo{pages}{3756--3775}.
\bibitem[{Hawking(1989)}]{hawking1989black}
\bibinfo{author}{S.~W. Hawking},
\newblock \bibinfo{title}{Black holes from cosmic strings},
\newblock \bibinfo{journal}{Physics Letters B} \bibinfo{volume}{231}
  (\bibinfo{year}{1989}) \bibinfo{pages}{237--239}.
\bibitem[{Polnarev and Zembowicz(1991)}]{polnarev1991formation}
\bibinfo{author}{A.~Polnarev}, \bibinfo{author}{R.~Zembowicz},
\newblock \bibinfo{title}{Formation of primordial black holes by cosmic
  strings},
\newblock \bibinfo{journal}{Physical Review D} \bibinfo{volume}{43}
  (\bibinfo{year}{1991}) \bibinfo{pages}{1106}.
\bibitem[{Garriga and Vilenkin(1993)}]{garriga1993black}
\bibinfo{author}{J.~Garriga}, \bibinfo{author}{A.~Vilenkin},
\newblock \bibinfo{title}{Black holes from nucleating strings},
\newblock \bibinfo{journal}{Physical Review D} \bibinfo{volume}{47}
  (\bibinfo{year}{1993}) \bibinfo{pages}{3265}.
\bibitem[{Vilenkin(1981)}]{vilenkin1981cosmological}
\bibinfo{author}{A.~Vilenkin},
\newblock \bibinfo{title}{Cosmological density fluctuations produced by vacuum
  strings},
\newblock \bibinfo{journal}{Physical Review Letters} \bibinfo{volume}{46}
  (\bibinfo{year}{1981}) \bibinfo{pages}{1169}.
\bibitem[{Fort and Vachaspati(1993)}]{fort1993global}
\bibinfo{author}{J.~Fort}, \bibinfo{author}{T.~Vachaspati},
\newblock \bibinfo{title}{Do global string loops collapse to form black
  holes?},
\newblock \bibinfo{journal}{Physics Letters B} \bibinfo{volume}{311}
  (\bibinfo{year}{1993}) \bibinfo{pages}{41--46}.
\bibitem[{Garriga and Sakellariadou(1993)}]{garriga1993effects}
\bibinfo{author}{J.~Garriga}, \bibinfo{author}{M.~Sakellariadou},
\newblock \bibinfo{title}{Effects of friction on cosmic strings},
\newblock \bibinfo{journal}{Physical Review D} \bibinfo{volume}{48}
  (\bibinfo{year}{1993}) \bibinfo{pages}{2502}.
\bibitem[{Rubin et~al.(2001)Rubin, Sakharov, and Khlopov}]{rubin2001formation}
\bibinfo{author}{S.~G. Rubin}, \bibinfo{author}{A.~S. Sakharov},
  \bibinfo{author}{M.~Y. Khlopov},
\newblock \bibinfo{title}{The formation of primary galactic nuclei during phase
  transitions in the early universe},
\newblock \bibinfo{journal}{Journal of Experimental and Theoretical Physics}
  \bibinfo{volume}{92} (\bibinfo{year}{2001}) \bibinfo{pages}{921--929}.
\bibitem[{Khlopov et~al.(2005)Khlopov, Rubin, and
  Sakharov}]{khlopov2005primordial}
\bibinfo{author}{M.~Y. Khlopov}, \bibinfo{author}{S.~G. Rubin},
  \bibinfo{author}{A.~S. Sakharov},
\newblock \bibinfo{title}{Primordial structure of massive black hole clusters},
\newblock \bibinfo{journal}{Astroparticle Physics} \bibinfo{volume}{23}
  (\bibinfo{year}{2005}) \bibinfo{pages}{265--277}.
\bibitem[{Garriga et~al.(2016)Garriga, Vilenkin, and Zhang}]{garriga2016black}
\bibinfo{author}{J.~Garriga}, \bibinfo{author}{A.~Vilenkin},
  \bibinfo{author}{J.~Zhang},
\newblock \bibinfo{title}{{Black holes and the multiverse}},
\newblock \bibinfo{journal}{JCAP} \bibinfo{volume}{1602} (\bibinfo{year}{2016})
  \bibinfo{pages}{064}.
\bibitem[{Deng et~al.(2017)Deng, Garriga, and Vilenkin}]{deng2017primordial}
\bibinfo{author}{H.~Deng}, \bibinfo{author}{J.~Garriga},
  \bibinfo{author}{A.~Vilenkin},
\newblock \bibinfo{title}{{Primordial black hole and wormhole formation by
  domain walls}},
\newblock \bibinfo{journal}{JCAP} \bibinfo{volume}{1704} (\bibinfo{year}{2017})
  \bibinfo{pages}{050}.
\bibitem[{Peccei and Quinn(1977)}]{peccei1977constraints}
\bibinfo{author}{R.~D. Peccei}, \bibinfo{author}{H.~R. Quinn},
\newblock \bibinfo{title}{Constraints imposed by cp conservation in the
  presence of pseudoparticles},
\newblock \bibinfo{journal}{Physical Review D} \bibinfo{volume}{16}
  (\bibinfo{year}{1977}) \bibinfo{pages}{1791}.
\bibitem[{Weinberg(1978)}]{weinberg1978new}
\bibinfo{author}{S.~Weinberg},
\newblock \bibinfo{title}{A new light boson?},
\newblock \bibinfo{journal}{Physical Review Letters} \bibinfo{volume}{40}
  (\bibinfo{year}{1978}) \bibinfo{pages}{223}.
\bibitem[{Wilczek(1978)}]{wilczek1978problem}
\bibinfo{author}{F.~Wilczek},
\newblock \bibinfo{title}{Problem of strong p and t invariance in the presence
  of instantons},
\newblock \bibinfo{journal}{Physical Review Letters} \bibinfo{volume}{40}
  (\bibinfo{year}{1978}) \bibinfo{pages}{279}.
\bibitem[{Kim(1979)}]{kim1979weak}
\bibinfo{author}{J.~E. Kim},
\newblock \bibinfo{title}{Weak-interaction singlet and strong cp invariance},
\newblock \bibinfo{journal}{Physical Review Letters} \bibinfo{volume}{43}
  (\bibinfo{year}{1979}) \bibinfo{pages}{103}.
\bibitem[{Shifman et~al.(1980)Shifman, Vainshtein, and
  Zakharov}]{shifman1980can}
\bibinfo{author}{M.~A. Shifman}, \bibinfo{author}{A.~Vainshtein},
  \bibinfo{author}{V.~I. Zakharov},
\newblock \bibinfo{title}{Can confinement ensure natural cp invariance of
  strong interactions?},
\newblock \bibinfo{journal}{Nuclear Physics B} \bibinfo{volume}{166}
  (\bibinfo{year}{1980}) \bibinfo{pages}{493--506}.
\bibitem[{Dine et~al.(1981)Dine, Fischler, and Srednicki}]{dine1981simple}
\bibinfo{author}{M.~Dine}, \bibinfo{author}{W.~Fischler},
  \bibinfo{author}{M.~Srednicki},
\newblock \bibinfo{title}{A simple solution to the strong cp problem with a
  harmless axion},
\newblock \bibinfo{journal}{Physics letters B} \bibinfo{volume}{104}
  (\bibinfo{year}{1981}) \bibinfo{pages}{199--202}.
\bibitem[{Zhitnitsky(1980)}]{Zhitnitsky:1980tq}
\bibinfo{author}{A.~R. Zhitnitsky},
\newblock \bibinfo{title}{{On Possible Suppression of the Axion Hadron
  Interactions. (In Russian)}},
\newblock \bibinfo{journal}{Sov. J. Nucl. Phys.} \bibinfo{volume}{31}
  (\bibinfo{year}{1980}) \bibinfo{pages}{260}. \bibinfo{note}{[Yad.
  Fiz.31,497(1980)]}.
\bibitem[{Vilenkin and Everett(1982)}]{vilenkin1982cosmic}
\bibinfo{author}{A.~Vilenkin}, \bibinfo{author}{A.~E. Everett},
\newblock \bibinfo{title}{Cosmic strings and domain walls in models with
  goldstone and pseudo-goldstone bosons},
\newblock \bibinfo{journal}{Physical Review Letters} \bibinfo{volume}{48}
  (\bibinfo{year}{1982}) \bibinfo{pages}{1867}.
\bibitem[{Sikivie(1982)}]{sikivie1982axions}
\bibinfo{author}{P.~Sikivie},
\newblock \bibinfo{title}{Axions, domain walls, and the early universe},
\newblock \bibinfo{journal}{Physical Review Letters} \bibinfo{volume}{48}
  (\bibinfo{year}{1982}) \bibinfo{pages}{1156}.
\bibitem[{Sikivie(2008)}]{sikivie2008axion}
\bibinfo{author}{P.~Sikivie},
\newblock \bibinfo{title}{Axion cosmology},
\newblock in: \bibinfo{booktitle}{Axions}, \bibinfo{publisher}{Springer},
  \bibinfo{year}{2008}, pp. \bibinfo{pages}{19--50}.
\bibitem[{Marsh(2016)}]{marsh2016axion}
\bibinfo{author}{D.~J. Marsh},
\newblock \bibinfo{title}{Axion cosmology},
\newblock \bibinfo{journal}{Physics Reports} \bibinfo{volume}{643}
  (\bibinfo{year}{2016}) \bibinfo{pages}{1--79}.
\bibitem[{Vachaspati(2017)}]{vachaspati2017lunar}
\bibinfo{author}{T.~Vachaspati},
\newblock \bibinfo{title}{Lunar mass black holes from qcd axion cosmology},
\newblock \bibinfo{journal}{arXiv preprint arXiv:1706.03868}
  (\bibinfo{year}{2017}).
\bibitem[{Ferrer et~al.(2019)Ferrer, Masso, Panico, Pujolas, and
  Rompineve}]{Ferrer:2018uiu}
\bibinfo{author}{F.~Ferrer}, \bibinfo{author}{E.~Masso},
  \bibinfo{author}{G.~Panico}, \bibinfo{author}{O.~Pujolas},
  \bibinfo{author}{F.~Rompineve},
\newblock \bibinfo{title}{Primordial black holes from the qcd axion},
\newblock \bibinfo{journal}{Physical review letters} \bibinfo{volume}{122}
  (\bibinfo{year}{2019}) \bibinfo{pages}{101301}.
\bibitem[{Hiramatsu et~al.(2012)Hiramatsu, Kawasaki, Saikawa, and
  Sekiguchi}]{hiramatsu2012production}
\bibinfo{author}{T.~Hiramatsu}, \bibinfo{author}{M.~Kawasaki},
  \bibinfo{author}{K.~Saikawa}, \bibinfo{author}{T.~Sekiguchi},
\newblock \bibinfo{title}{Production of dark matter axions from collapse of
  string-wall systems},
\newblock \bibinfo{journal}{Physical Review D} \bibinfo{volume}{85}
  (\bibinfo{year}{2012}) \bibinfo{pages}{105020}.
\bibitem[{Fleury and Moore(2016)}]{Fleury:2015aca}
\bibinfo{author}{L.~Fleury}, \bibinfo{author}{G.~D. Moore},
\newblock \bibinfo{title}{{Axion dark matter: strings and their cores}},
\newblock \bibinfo{journal}{JCAP} \bibinfo{volume}{1601} (\bibinfo{year}{2016})
  \bibinfo{pages}{004}.
\bibitem[{Klaer and Moore(2017)}]{Klaer:2017ond}
\bibinfo{author}{V.~B. Klaer}, \bibinfo{author}{G.~D. Moore},
\newblock \bibinfo{title}{{The dark-matter axion mass}},
\newblock \bibinfo{journal}{JCAP} \bibinfo{volume}{1711} (\bibinfo{year}{2017})
  \bibinfo{pages}{049}.
\bibitem[{Gorghetto et~al.(2018)Gorghetto, Hardy, and
  Villadoro}]{Gorghetto:2018myk}
\bibinfo{author}{M.~Gorghetto}, \bibinfo{author}{E.~Hardy},
  \bibinfo{author}{G.~Villadoro},
\newblock \bibinfo{title}{{Axions from Strings: the Attractive Solution}},
\newblock \bibinfo{journal}{JHEP} \bibinfo{volume}{07} (\bibinfo{year}{2018})
  \bibinfo{pages}{151}.
\bibitem[{Kawasaki et~al.(2018)Kawasaki, Sekiguchi, Yamaguchi, and
  Yokoyama}]{Kawasaki:2018bzv}
\bibinfo{author}{M.~Kawasaki}, \bibinfo{author}{T.~Sekiguchi},
  \bibinfo{author}{M.~Yamaguchi}, \bibinfo{author}{J.~Yokoyama},
\newblock \bibinfo{title}{{Long-term dynamics of cosmological axion strings}},
\newblock \bibinfo{journal}{PTEP} \bibinfo{volume}{2018} (\bibinfo{year}{2018})
  \bibinfo{pages}{091E01}.
\bibitem[{Zeldovich et~al.(1974)Zeldovich, Kobzarev, and
  Okun}]{Zeldovich:1974uw}
\bibinfo{author}{{\relax Ya}.~B. Zeldovich}, \bibinfo{author}{I.~{\relax Yu}.
  Kobzarev}, \bibinfo{author}{L.~B. Okun},
\newblock \bibinfo{title}{{Cosmological Consequences of the Spontaneous
  Breakdown of Discrete Symmetry}},
\newblock \bibinfo{journal}{Zh. Eksp. Teor. Fiz.} \bibinfo{volume}{67}
  (\bibinfo{year}{1974}) \bibinfo{pages}{3--11}. \bibinfo{note}{[Sov. Phys.
  JETP40,1(1974)]}.
\bibitem[{Peccei(2008)}]{peccei2008strong}
\bibinfo{author}{R.~D. Peccei},
\newblock \bibinfo{title}{The strong cp problem and axions},
\newblock in: \bibinfo{booktitle}{Axions}, \bibinfo{publisher}{Springer},
  \bibinfo{year}{2008}, pp. \bibinfo{pages}{3--17}.
\bibitem[{Borsanyi et~al.(2016)}]{Borsanyi:2016ksw}
\bibinfo{author}{S.~Borsanyi}, et~al.,
\newblock \bibinfo{title}{{Calculation of the axion mass based on
  high-temperature lattice quantum chromodynamics}},
\newblock \bibinfo{journal}{Nature} \bibinfo{volume}{539}
  (\bibinfo{year}{2016}) \bibinfo{pages}{69--71}.
\bibitem[{Wantz and Shellard(2010)}]{wantz2010axion}
\bibinfo{author}{O.~Wantz}, \bibinfo{author}{E.~Shellard},
\newblock \bibinfo{title}{Axion cosmology revisited},
\newblock \bibinfo{journal}{Physical Review D} \bibinfo{volume}{82}
  (\bibinfo{year}{2010}) \bibinfo{pages}{123508}.
\bibitem[{Gorghetto and Villadoro(2019)}]{Gorghetto:2018ocs}
\bibinfo{author}{M.~Gorghetto}, \bibinfo{author}{G.~Villadoro},
\newblock \bibinfo{title}{{Topological Susceptibility and QCD Axion Mass: QED
  and NNLO corrections}},
\newblock \bibinfo{journal}{JHEP} \bibinfo{volume}{03} (\bibinfo{year}{2019})
  \bibinfo{pages}{033}.
\bibitem[{Kibble(1976)}]{Kibble:1976sj}
\bibinfo{author}{T.~W.~B. Kibble},
\newblock \bibinfo{title}{{Topology of Cosmic Domains and Strings}},
\newblock \bibinfo{journal}{J. Phys.} \bibinfo{volume}{A9}
  (\bibinfo{year}{1976}) \bibinfo{pages}{1387--1398}.
\bibitem[{Zurek(1985)}]{Zurek:1985qw}
\bibinfo{author}{W.~H. Zurek},
\newblock \bibinfo{title}{{Cosmological Experiments in Superfluid Helium?}},
\newblock \bibinfo{journal}{Nature} \bibinfo{volume}{317}
  (\bibinfo{year}{1985}) \bibinfo{pages}{505--508}.
\bibitem[{Liang and Zhitnitsky(2016)}]{Liang:2016tqc}
\bibinfo{author}{X.~Liang}, \bibinfo{author}{A.~Zhitnitsky},
\newblock \bibinfo{title}{{Axion field and the quark nugget's formation at the
  QCD phase transition}},
\newblock \bibinfo{journal}{Phys. Rev.} \bibinfo{volume}{D94}
  (\bibinfo{year}{2016}) \bibinfo{pages}{083502}.
\bibitem[{Vachaspati(2006)}]{vachaspati2006kinks}
\bibinfo{author}{T.~Vachaspati}, \bibinfo{title}{Kinks and domain walls: An
  introduction to classical and quantum solitons},
  \bibinfo{publisher}{Cambridge University Press}, \bibinfo{year}{2006}.
\bibitem[{Vilenkin and Shellard(2000)}]{Vilenkin:2000jqa}
\bibinfo{author}{A.~Vilenkin}, \bibinfo{author}{E.~P.~S. Shellard},
  \bibinfo{title}{{Cosmic Strings and Other Topological Defects}},
  \bibinfo{publisher}{Cambridge University Press}, \bibinfo{year}{2000}.
\bibitem[{Forbes and Zhitnitsky(2001)}]{forbes2001domain}
\bibinfo{author}{M.~M. Forbes}, \bibinfo{author}{A.~R. Zhitnitsky},
\newblock \bibinfo{title}{Domain walls in qcd},
\newblock \bibinfo{journal}{Journal of High Energy Physics}
  \bibinfo{volume}{2001} (\bibinfo{year}{2001}) \bibinfo{pages}{013}.
\bibitem[{Ge et~al.(2019)Ge, Lawson, and Zhitnitsky}]{Ge:2019voa}
\bibinfo{author}{S.~Ge}, \bibinfo{author}{K.~Lawson},
  \bibinfo{author}{A.~Zhitnitsky},
\newblock \bibinfo{title}{{Axion quark nugget dark matter model: Size
  distribution and survival pattern}},
\newblock \bibinfo{journal}{Phys. Rev.} \bibinfo{volume}{D99}
  (\bibinfo{year}{2019}) \bibinfo{pages}{116017}.
\bibitem[{Vachaspati and Vilenkin(1984)}]{vachaspati1984formation}
\bibinfo{author}{T.~Vachaspati}, \bibinfo{author}{A.~Vilenkin},
\newblock \bibinfo{title}{Formation and evolution of cosmic strings},
\newblock \bibinfo{journal}{Physical Review D} \bibinfo{volume}{30}
  (\bibinfo{year}{1984}) \bibinfo{pages}{2036}.
\bibitem[{Harvey et~al.(1982)Harvey, Kolb, Reiss, and
  Wolfram}]{harvey1982calculation}
\bibinfo{author}{J.~A. Harvey}, \bibinfo{author}{E.~W. Kolb},
  \bibinfo{author}{D.~B. Reiss}, \bibinfo{author}{S.~Wolfram},
\newblock \bibinfo{title}{Calculation of cosmological baryon asymmetry in grand
  unified gauge models},
\newblock \bibinfo{journal}{Nuclear Physics B} \bibinfo{volume}{201}
  (\bibinfo{year}{1982}) \bibinfo{pages}{16--100}.
\bibitem[{Stauffer(1979)}]{STAUFFER19791}
\bibinfo{author}{D.~Stauffer},
\newblock \bibinfo{title}{Scaling theory of percolation clusters},
\newblock \bibinfo{journal}{Physics Reports} \bibinfo{volume}{54}
  (\bibinfo{year}{1979}) \bibinfo{pages}{1 -- 74}.
\bibitem[{Isichenko(1992)}]{isichenko1992percolation}
\bibinfo{author}{M.~B. Isichenko},
\newblock \bibinfo{title}{Percolation, statistical topography, and transport in
  random media},
\newblock \bibinfo{journal}{Reviews of modern physics} \bibinfo{volume}{64}
  (\bibinfo{year}{1992}) \bibinfo{pages}{961}.
\bibitem[{Grinchuk(2002)}]{grinchuk2002large}
\bibinfo{author}{P.~Grinchuk},
\newblock \bibinfo{title}{Large clusters in supercritical percolation},
\newblock \bibinfo{journal}{Physical Review E} \bibinfo{volume}{66}
  (\bibinfo{year}{2002}) \bibinfo{pages}{016124}.
\bibitem[{Stauffer and Aharony(2014)}]{stauffer2014introduction}
\bibinfo{author}{D.~Stauffer}, \bibinfo{author}{A.~Aharony},
  \bibinfo{title}{Introduction to percolation theory: revised second edition},
  \bibinfo{publisher}{CRC press}, \bibinfo{year}{2014}.
\bibitem[{Lubensky and McKane(1981)}]{lubensky1981cluster}
\bibinfo{author}{T.~Lubensky}, \bibinfo{author}{A.~McKane},
\newblock \bibinfo{title}{Cluster size distribution above the percolation
  threshold},
\newblock \bibinfo{journal}{Journal of Physics A: Mathematical and General}
  \bibinfo{volume}{14} (\bibinfo{year}{1981}) \bibinfo{pages}{L157}.
\bibitem[{Bauchspiess and Stauffer(1978)}]{BAUCHSPIESS1978567}
\bibinfo{author}{K.~Bauchspiess}, \bibinfo{author}{D.~Stauffer},
\newblock \bibinfo{title}{Use of percolation clusters in nucleation theory},
\newblock \bibinfo{journal}{Journal of Aerosol Science} \bibinfo{volume}{9}
  (\bibinfo{year}{1978}) \bibinfo{pages}{567 -- 577}.
\bibitem[{Chang et~al.(1998)Chang, Hagmann, and Sikivie}]{Chang:1998tb}
\bibinfo{author}{S.~Chang}, \bibinfo{author}{C.~Hagmann},
  \bibinfo{author}{P.~Sikivie},
\newblock \bibinfo{title}{{Studies of the motion and decay of axion walls
  bounded by strings}},
\newblock \bibinfo{journal}{Phys. Rev.} \bibinfo{volume}{D59}
  (\bibinfo{year}{1998}) \bibinfo{pages}{023505}.
\bibitem[{Zhitnitsky(2003)}]{Zhitnitsky:2002qa}
\bibinfo{author}{A.~R. Zhitnitsky},
\newblock \bibinfo{title}{{'Nonbaryonic' dark matter as baryonic color
  superconductor}},
\newblock \bibinfo{journal}{JCAP} \bibinfo{volume}{0310} (\bibinfo{year}{2003})
  \bibinfo{pages}{010}.
\bibitem[{Ge et~al.(2017)Ge, Liang, and Zhitnitsky}]{Ge:2017ttc}
\bibinfo{author}{S.~Ge}, \bibinfo{author}{X.~Liang},
  \bibinfo{author}{A.~Zhitnitsky},
\newblock \bibinfo{title}{{Cosmological CP odd axion field as the coherent
  Berry's phase of the Universe}},
\newblock \bibinfo{journal}{Phys. Rev.} \bibinfo{volume}{D96}
  (\bibinfo{year}{2017}) \bibinfo{pages}{063514}.
\bibitem[{Ge et~al.(2018)Ge, Liang, and Zhitnitsky}]{Ge:2017_2}
\bibinfo{author}{S.~Ge}, \bibinfo{author}{X.~Liang},
  \bibinfo{author}{A.~Zhitnitsky},
\newblock \bibinfo{title}{{Cosmological axion and a quark nugget dark matter
  model}},
\newblock \bibinfo{journal}{Phys. Rev.} \bibinfo{volume}{D97}
  (\bibinfo{year}{2018}) \bibinfo{pages}{043008}.
\bibitem[{Zhitnitsky(2017)}]{Zhitnitsky:2017rop}
\bibinfo{author}{A.~Zhitnitsky},
\newblock \bibinfo{title}{{Solar Extreme UV radiation and quark nugget dark
  matter model}},
\newblock \bibinfo{journal}{JCAP} \bibinfo{volume}{1710} (\bibinfo{year}{2017})
  \bibinfo{pages}{050}.
\bibitem[{Lawson and Zhitnitsky(2018)}]{Lawson:2018qkc}
\bibinfo{author}{K.~Lawson}, \bibinfo{author}{A.~R. Zhitnitsky},
\newblock \bibinfo{title}{{The 21 cm absorption line and the axion quark nugget
  dark matter model}},
\newblock \bibinfo{journal}{Phys. Dark Univ.}  (\bibinfo{year}{2018})
  \bibinfo{pages}{100295}. \bibinfo{note}{[Phys. Dark Univ.100295,2019(2018)]}.
\bibitem[{Raza et~al.(2018)Raza, van Waerbeke, and Zhitnitsky}]{Raza:2018gpb}
\bibinfo{author}{N.~Raza}, \bibinfo{author}{L.~van Waerbeke},
  \bibinfo{author}{A.~Zhitnitsky},
\newblock \bibinfo{title}{{Solar corona heating by axion quark nugget dark
  matter}},
\newblock \bibinfo{journal}{Phys. Rev.} \bibinfo{volume}{D98}
  (\bibinfo{year}{2018}) \bibinfo{pages}{103527}.
\bibitem[{Fischer et~al.(2018)Fischer, Liang, Semertzidis, Zhitnitsky, and
  Zioutas}]{Fischer:2018niu}
\bibinfo{author}{H.~Fischer}, \bibinfo{author}{X.~Liang},
  \bibinfo{author}{Y.~Semertzidis}, \bibinfo{author}{A.~Zhitnitsky},
  \bibinfo{author}{K.~Zioutas},
\newblock \bibinfo{title}{{New mechanism producing axions in the AQN model and
  how the CAST can discover them}},
\newblock \bibinfo{journal}{Phys. Rev.} \bibinfo{volume}{D98}
  (\bibinfo{year}{2018}) \bibinfo{pages}{043013}.
\bibitem[{van Waerbeke and Zhitnitsky(2019)}]{vanWaerbeke:2018nyj}
\bibinfo{author}{L.~van Waerbeke}, \bibinfo{author}{A.~Zhitnitsky},
\newblock \bibinfo{title}{{Fast Radio Bursts and the Axion Quark Nugget Dark
  Matter Model}},
\newblock \bibinfo{journal}{Phys. Rev.} \bibinfo{volume}{D99}
  (\bibinfo{year}{2019}) \bibinfo{pages}{043535}.
\bibitem[{Liang and Zhitnitsky(2019)}]{Liang:2018ecs}
\bibinfo{author}{X.~Liang}, \bibinfo{author}{A.~Zhitnitsky},
\newblock \bibinfo{title}{{Gravitationally bound axions and how one can
  discover them}},
\newblock \bibinfo{journal}{Phys. Rev.} \bibinfo{volume}{D99}
  (\bibinfo{year}{2019}) \bibinfo{pages}{023015}.
\bibitem[{Flambaum and Zhitnitsky(2019)}]{Flambaum:2018ohm}
\bibinfo{author}{V.~V. Flambaum}, \bibinfo{author}{A.~R. Zhitnitsky},
\newblock \bibinfo{title}{{Primordial Lithium Puzzle and the Axion Quark Nugget
  Dark Matter Model}},
\newblock \bibinfo{journal}{Phys. Rev.} \bibinfo{volume}{D99}
  (\bibinfo{year}{2019}) \bibinfo{pages}{023517}.
\bibitem[{Lawson et~al.(2019)Lawson, Liang, Mead, Siddiqui, Van~Waerbeke, and
  Zhitnitsky}]{Lawson:2019cvy}
\bibinfo{author}{K.~Lawson}, \bibinfo{author}{X.~Liang},
  \bibinfo{author}{A.~Mead}, \bibinfo{author}{M.~S.~R. Siddiqui},
  \bibinfo{author}{L.~Van~Waerbeke}, \bibinfo{author}{A.~Zhitnitsky},
\newblock \bibinfo{title}{{Gravitationally trapped axions on the Earth}},
\newblock \bibinfo{journal}{Phys. Rev.} \bibinfo{volume}{D100}
  (\bibinfo{year}{2019}) \bibinfo{pages}{043531}.
\bibitem[{Graham et~al.(2015)Graham, Irastorza, Lamoreaux, Lindner, and van
  Bibber}]{graham2015experimental}
\bibinfo{author}{P.~W. Graham}, \bibinfo{author}{I.~G. Irastorza},
  \bibinfo{author}{S.~K. Lamoreaux}, \bibinfo{author}{A.~Lindner},
  \bibinfo{author}{K.~A. van Bibber},
\newblock \bibinfo{title}{Experimental searches for the axion and axion-like
  particles},
\newblock \bibinfo{journal}{Annual Review of Nuclear and Particle Science}
  \bibinfo{volume}{65} (\bibinfo{year}{2015}) \bibinfo{pages}{485--514}.
\bibitem[{Barnacka et~al.(2012)Barnacka, Glicenstein, and
  Moderski}]{femtolensing}
\bibinfo{author}{A.~Barnacka}, \bibinfo{author}{J.-F. Glicenstein},
  \bibinfo{author}{R.~Moderski},
\newblock \bibinfo{title}{New constraints on primordial black holes abundance
  from femtolensing of gamma-ray bursts},
\newblock \bibinfo{journal}{Physical Review D} \bibinfo{volume}{86}
  (\bibinfo{year}{2012}) \bibinfo{pages}{043001}.
\bibitem[{Graham et~al.(2015)Graham, Rajendran, and Varela}]{WD}
\bibinfo{author}{P.~W. Graham}, \bibinfo{author}{S.~Rajendran},
  \bibinfo{author}{J.~Varela},
\newblock \bibinfo{title}{Dark matter triggers of supernovae},
\newblock \bibinfo{journal}{Physical Review D} \bibinfo{volume}{92}
  (\bibinfo{year}{2015}) \bibinfo{pages}{063007}.
\bibitem[{Niikura et~al.(2019)Niikura, Takada, Yasuda, Lupton, Sumi, More,
  Kurita, Sugiyama, More, Oguri et~al.}]{HSCmicrolensing}
\bibinfo{author}{H.~Niikura}, \bibinfo{author}{M.~Takada},
  \bibinfo{author}{N.~Yasuda}, \bibinfo{author}{R.~H. Lupton},
  \bibinfo{author}{T.~Sumi}, \bibinfo{author}{S.~More},
  \bibinfo{author}{T.~Kurita}, \bibinfo{author}{S.~Sugiyama},
  \bibinfo{author}{A.~More}, \bibinfo{author}{M.~Oguri}, et~al.,
\newblock \bibinfo{title}{Microlensing constraints on primordial black holes
  with subaru/hsc andromeda observations},
\newblock \bibinfo{journal}{Nature Astronomy}  (\bibinfo{year}{2019})
  \bibinfo{pages}{1}.
\bibitem[{Griest et~al.(2014)Griest, Cieplak, and Lehner}]{Keplermicrolensing}
\bibinfo{author}{K.~Griest}, \bibinfo{author}{A.~M. Cieplak},
  \bibinfo{author}{M.~J. Lehner},
\newblock \bibinfo{title}{Experimental limits on primordial black hole dark
  matter from the first 2 yr of kepler data},
\newblock \bibinfo{journal}{The Astrophysical Journal} \bibinfo{volume}{786}
  (\bibinfo{year}{2014}) \bibinfo{pages}{158}.
\bibitem[{Fuller et~al.(2017)Fuller, Kusenko, and
  Takhistov}]{fuller2017primordial}
\bibinfo{author}{G.~M. Fuller}, \bibinfo{author}{A.~Kusenko},
  \bibinfo{author}{V.~Takhistov},
\newblock \bibinfo{title}{Primordial black holes and r-process
  nucleosynthesis},
\newblock \bibinfo{journal}{Physical review letters} \bibinfo{volume}{119}
  (\bibinfo{year}{2017}) \bibinfo{pages}{061101}.
\bibitem[{Capela et~al.(2013)Capela, Pshirkov, and Tinyakov}]{Capela:2013yf}
\bibinfo{author}{F.~Capela}, \bibinfo{author}{M.~Pshirkov},
  \bibinfo{author}{P.~Tinyakov},
\newblock \bibinfo{title}{{Constraints on primordial black holes as dark matter
  candidates from capture by neutron stars}},
\newblock \bibinfo{journal}{Phys. Rev.} \bibinfo{volume}{D87}
  (\bibinfo{year}{2013}) \bibinfo{pages}{123524}.
\bibitem[{Lane et~al.(2009)Lane, Kiss, Lewis, Ibata, Siebert, Bedding, and
  Sz{\'e}kely}]{lane2009testing}
\bibinfo{author}{R.~R. Lane}, \bibinfo{author}{L.~L. Kiss},
  \bibinfo{author}{G.~F. Lewis}, \bibinfo{author}{R.~A. Ibata},
  \bibinfo{author}{A.~Siebert}, \bibinfo{author}{T.~R. Bedding},
  \bibinfo{author}{P.~Sz{\'e}kely},
\newblock \bibinfo{title}{Testing newtonian gravity with aaomega: mass-to-light
  profiles of four globular clusters},
\newblock \bibinfo{journal}{Monthly Notices of the Royal Astronomical Society}
  \bibinfo{volume}{400} (\bibinfo{year}{2009}) \bibinfo{pages}{917--923}.
\bibitem[{Chang et~al.(2018)Chang, Essig, and McDermott}]{chang2018supernova}
\bibinfo{author}{J.~H. Chang}, \bibinfo{author}{R.~Essig},
  \bibinfo{author}{S.~D. McDermott},
\newblock \bibinfo{title}{Supernova 1987a constraints on sub-gev dark sectors,
  millicharged particles, the qcd axion, and an axion-like particle},
\newblock \bibinfo{journal}{Journal of High Energy Physics}
  \bibinfo{volume}{2018} (\bibinfo{year}{2018}) \bibinfo{pages}{51}.
\bibitem[{Ringwald(2014)}]{Ringwald:2014vqa}
\bibinfo{author}{A.~Ringwald},
\newblock \bibinfo{title}{{Axions and Axion-Like Particles}},
\newblock in: \bibinfo{booktitle}{{Proceedings, 49th Rencontres de Moriond on
  Electroweak Interactions and Unified Theories: La Thuile, Italy, March 15-22,
  2014}}, pp. \bibinfo{pages}{223--230}.
\bibitem[{Irastorza and Redondo(2018)}]{Irastorza:2018dyq}
\bibinfo{author}{I.~G. Irastorza}, \bibinfo{author}{J.~Redondo},
\newblock \bibinfo{title}{{New experimental approaches in the search for
  axion-like particles}},
\newblock \bibinfo{journal}{Prog. Part. Nucl. Phys.} \bibinfo{volume}{102}
  (\bibinfo{year}{2018}) \bibinfo{pages}{89--159}.
\bibitem[{Côté et~al.(2017)Côté, Belczynski, Fryer, Ritter, Paul, Wehmeyer,
  and O'Shea}]{Cote:2016vla}
\bibinfo{author}{B.~Côté}, \bibinfo{author}{K.~Belczynski},
  \bibinfo{author}{C.~L. Fryer}, \bibinfo{author}{C.~Ritter},
  \bibinfo{author}{A.~Paul}, \bibinfo{author}{B.~Wehmeyer},
  \bibinfo{author}{B.~W. O'Shea},
\newblock \bibinfo{title}{{Advanced LIGO Constraints on Neutron Star Mergers
  and R-Process Sites}},
\newblock \bibinfo{journal}{Astrophys. J.} \bibinfo{volume}{836}
  (\bibinfo{year}{2017}) \bibinfo{pages}{230}.
\bibitem[{Acernese et~al.(2015)}]{TheVirgo:2014hva}
\bibinfo{author}{F.~Acernese}, et~al.,
\newblock \bibinfo{title}{{Advanced Virgo: a second-generation interferometric
  gravitational wave detector}},
\newblock \bibinfo{journal}{Class. Quant. Grav.} \bibinfo{volume}{32}
  (\bibinfo{year}{2015}) \bibinfo{pages}{024001}.
\bibitem[{Aso et~al.(2013)Aso, Michimura, Somiya, Ando, Miyakawa, Sekiguchi,
  Tatsumi, and Yamamoto}]{Aso:2013eba}
\bibinfo{author}{Y.~Aso}, \bibinfo{author}{Y.~Michimura},
  \bibinfo{author}{K.~Somiya}, \bibinfo{author}{M.~Ando},
  \bibinfo{author}{O.~Miyakawa}, \bibinfo{author}{T.~Sekiguchi},
  \bibinfo{author}{D.~Tatsumi}, \bibinfo{author}{H.~Yamamoto},
\newblock \bibinfo{title}{{Interferometer design of the KAGRA gravitational
  wave detector}},
\newblock \bibinfo{journal}{Phys. Rev.} \bibinfo{volume}{D88}
  (\bibinfo{year}{2013}) \bibinfo{pages}{043007}.
\bibitem[{Inoue and Tanaka(2003)}]{PhysRevLett.91.021101}
\bibinfo{author}{K.~T. Inoue}, \bibinfo{author}{T.~Tanaka},
\newblock \bibinfo{title}{Gravitational waves from sub-lunar-mass primordial
  black-hole binaries: A new probe of extradimensions},
\newblock \bibinfo{journal}{Phys. Rev. Lett.} \bibinfo{volume}{91}
  (\bibinfo{year}{2003}) \bibinfo{pages}{021101}.
\bibitem[{Naderi et~al.(2018)Naderi, Mehrabi, and Rahvar}]{Naderi:2017ite}
\bibinfo{author}{T.~Naderi}, \bibinfo{author}{A.~Mehrabi},
  \bibinfo{author}{S.~Rahvar},
\newblock \bibinfo{title}{{Primordial black hole detection through diffractive
  microlensing}},
\newblock \bibinfo{journal}{Phys. Rev.} \bibinfo{volume}{D97}
  (\bibinfo{year}{2018}) \bibinfo{pages}{103507}.
\bibitem[{Vilenkin(1985)}]{Vilenkin:1984ib}
\bibinfo{author}{A.~Vilenkin},
\newblock \bibinfo{title}{{Cosmic Strings and Domain Walls}},
\newblock \bibinfo{journal}{Phys. Rept.} \bibinfo{volume}{121}
  (\bibinfo{year}{1985}) \bibinfo{pages}{263--315}.
\bibitem[{{Shellard}(1986)}]{Shellard}
\bibinfo{author}{E.~P.~S. {Shellard}},
\newblock \bibinfo{title}{{Axionic domain walls and cosmology}},
\newblock in: \bibinfo{booktitle}{Liege International Astrophysical Colloquia},
  volume~\bibinfo{volume}{26} of \textit{\bibinfo{series}{Liege International
  Astrophysical Colloquia}}, pp. \bibinfo{pages}{173--179}.
\bibitem[{{Ryden} et~al.(1990){Ryden}, {Press}, and {Spergel}}]{Ryden}
\bibinfo{author}{B.~S. {Ryden}}, \bibinfo{author}{W.~H. {Press}},
  \bibinfo{author}{D.~N. {Spergel}},
\newblock \bibinfo{title}{{The evolution of networks of domain walls and cosmic
  strings}},
\newblock \bibinfo{journal}{Astrophys. J} \bibinfo{volume}{357}
  (\bibinfo{year}{1990}) \bibinfo{pages}{293--300}.
\bibitem[{{Press} et~al.(1989){Press}, {Ryden}, and {Spergel}}]{Press}
\bibinfo{author}{W.~H. {Press}}, \bibinfo{author}{B.~S. {Ryden}},
  \bibinfo{author}{D.~N. {Spergel}},
\newblock \bibinfo{title}{{Dynamical evolution of domain walls in an expanding
  universe}},
\newblock \bibinfo{journal}{Astrophys. J} \bibinfo{volume}{347}
  (\bibinfo{year}{1989}) \bibinfo{pages}{590--604}.

\end{thebibliography}

\end{document}